\documentclass[12pt,a4paper]{article}

\usepackage[margin=2.5cm]{geometry}
\usepackage{amsmath,amssymb}
\usepackage{bm}
\usepackage{booktabs}
\usepackage{array}
\usepackage{longtable}
\usepackage{hyperref}
\usepackage{xcolor}
\usepackage{graphicx}
\usepackage{caption}
\usepackage{subcaption}
\usepackage{enumitem}
\usepackage{natbib}
\usepackage{microtype}
\usepackage{setspace}
\usepackage{titlesec}
\usepackage{abstract}

\newcommand{\MBH}{M_{\bullet}}
\newcommand{\Msun}{M_{\odot}}
\newcommand{\Mdot}{\dot{M}}
\newcommand{\mdot}{\dot{m}}
\newcommand{\astar}{a_{\star}}
\newcommand{\SgrA}{Sgr~A$^{\star}$}
\newcommand{\tact}{T_{\mathrm{act}}}
\newcommand{\Myr}{\mathrm{Myr}}

\newcommand{\erg}{\mathrm{erg}}
\newcommand{\ergs}{\mathrm{erg\,s^{-1}}}
\newcommand{\BZ}{\mathrm{BZ}}
\newcommand{\Edd}{\mathrm{Edd}}

\newcommand{\CR}{\mathrm{CR}}

\newcommand{\GCE}{\mathrm{GCE}}
\newcommand{\kpc}{\,\mathrm{kpc}}

\newcommand{\Prec}{\mathcal{N}_{\mathrm{prec}}}
\newcommand{\erfc}{\operatorname{erfc}}

\newcommand{\wall}{\mathrm{wall}}
\newcommand{\tilt}{\mathrm{tilt}}
\newcommand{\GeV}{\mathrm{GeV}}
\newcommand{\TeV}{\mathrm{TeV}}

\newcommand{\PeV}{\mathrm{PeV}}
 
\usepackage{xcolor}
\usepackage[dvipsnames]{xcolor}

\hypersetup{
  colorlinks = true,
  linkcolor  = blue!70!black,
  citecolor  = blue!70!black,
  urlcolor   = blue!70!black,
  pdfauthor  = {}
}

\titleformat{\section}{\large\bfseries}{\thesection.}{0.6em}{}[\vspace{-2pt}\rule{\linewidth}{0.4pt}]
\titleformat{\subsection}{\normalsize\bfseries}{\thesubsection}{0.6em}{}
\titleformat{\subsubsection}{\normalsize\itshape}{\thesubsubsection}{0.6em}{}

\begin{document}

\begin{titlepage}

\begin{center}
  \vspace*{1cm}
  {\LARGE\bfseries
     Precessing Black Hole Jets \\[6pt]
     and \\[6pt]   \vspace{0.3cm}Galactic Fossils
     }
  \vspace{1.cm}

  {\large Mar\'ia J. Rodr\'iguez\textsuperscript{1,2,*}}\\[0.8em]
  \vspace*{0.3cm}
  {\small
    \textsuperscript{1}\textit{Department of Physics, Utah State University,\\
    4415 Old Main Hill Road, Logan, UT 84322, USA}\\[0.3em]
    \textsuperscript{2}\textit{Instituto de F\'isica Te\'orica UAM--CSIC,\\
    Universidad Aut\'onoma de Madrid, 28049 Madrid, Spain}\\[0.6em]
    \textsuperscript{*}\href{mailto:majo.rodriguez.b@gmail.com}{majo.rodriguez.b@gmail.com}
  }

  \vspace{1.cm}
  \noindent\rule{0.85\linewidth}{0.5pt}

\noindent
\begin{abstract}
The Galactic Center gamma-ray excess (GCE)---a
few-GeV excess in the Fermi-LAT data ---has remained without
a consensus interpretation for more than fifteen years.
Dark-matter annihilation and unresolved millisecond-pulsar
populations remain the leading candidates, yet neither
connects the excess to the past activity of \SgrA{}
traced by the Fermi and eROSITA bubbles. We propose a
common-origin scenario in which a contribution to the GCE
arises as a fossil hadronic imprint of the same \SgrA{}
outburst associated with the bubbles. We develop a model of
\SgrA{} with a precessing paraboloidal Blandford--Znajek (BZ) jet
launched from a tilted, magnetically arrested accretion disc
during a $\sim 7.5\,\Myr$ active phase ending
$\sim 2.6\,\Myr$ ago. In this picture, the jet drives the
bipolar expansion of the Fermi/eROSITA
bubbles---contributing, alongside wider-angle outflows, to
their observed extent---and injects hadronic cosmic rays
at the Galactic Center. We couple the analytic BZ injection
to a two-zone diffusion numerical solver to compute the
resulting GCE surface brightness and verify internal
consistency: the proton Larmor radius remains small compared
to the jet coherence scale, ensuring magnetic confinement of
the cosmic-ray population, while attenuation of the produced
$\gamma$ rays in the interstellar medium is negligible,
leaving the medium effectively transparent. Isolating the
jet contribution alone yields a spin-dependent, irreducible
hadronic floor: for a \SgrA{} spin of $a_\star = 0.9$, we
find a robust floor at the few-percent to $\sim 10\%$ level
of the observed GCE surface brightness across the inner ten
degrees, highlighting a previously unexplored component
relevant for comprehensive models of the GCE.
\end{abstract}

\noindent\rule{0.85\linewidth}{0.5pt}

\end{center}
\end{titlepage}

\tableofcontents
\clearpage

\section{Introduction}
\label{sec:intro}
The \textit{Fermi} Large Area Telescope has revealed an
extended GeV gamma-ray excess centered on the Galactic Center,
detected out to $\sim 10^\circ$ ($\sim 1.5\,\mathrm{kpc}$
projected) from Sagittarius A* and exhibiting an
approximately spherically symmetric
morphology~\cite{Goodenough2009,Hooper2011,Gordon2013,Daylan2016}.
This Galactic Center Excess (GCE) is a few-GeV gamma-ray excess with a spectrum peaking near
$E_\gamma \simeq 2\,\mathrm{GeV}$ and a morphology compatible
with a squared generalized NFW profile with inner slope
$\gamma \simeq 1.2$–$1.3$~\cite{DiMauro2021,Calore2015},
making it one of the most widely studied anomalies in
high-energy astrophysics.

Two primary interpretations have dominated the literature:
the dark-matter (DM) annihilation
hypothesis~\cite{Goodenough2009,Hooper2011,Daylan2016,DiMauro2021},
attributing the excess to $\sim 10$--$100\,\GeV$ WIMPs
annihilating to hadronic final states, with $b\bar{b}$
providing a representative fit, and the unresolved
millisecond-pulsar (MSP)
hypothesis~\cite{Lee2016,Bartels2016,HolstHooper2023,Sands2025b,Kalambay2026},
which notes that the GCE spectrum is spectrally consistent
with a superposition of MSP-like sources associated with the
bulge stellar population. Both interpretations face
challenges: the DM scenario is in tension with expectations
from baryonic effects and bar-induced modifications of the
inner halo profile, while the MSP interpretation requires a
population that is somewhat fainter and more numerous than
standard Galactic MSP scaling relations, subject to current
point-source sensitivity constraints. A third class of
mechanisms, less developed in the GCE literature, invokes
hadronic CRs from past Galactic-center activity --- including
supernovae, stellar winds, and Super Massive Black Hole (SMBH) outbursts; the present
work develops a specific realisation in the SMBH-outburst
class.

A fundamental limitation common to DM and MSP
interpretations is that neither incorporates the past
activity of \SgrA{} directly traced by the Fermi and eROSITA
bubbles. These giant $\gamma$-ray and X-ray lobes, extending
$\sim 8$--$14\,\mathrm{kpc}$ above and below the Galactic
plane and highly symmetric about the disc and rotation
axis~\cite{Su2010Fermi,Predehl2020,Ackermann2014bubbles},
provide strong evidence for past energetic activity in the
Galactic Center on Myr timescales. Cosmological simulations
indicate that such structures are common in disc galaxies of
similar mass: a large fraction of Milky Way/M31 analogues in
TNG50 host comparable circumgalactic X-ray bubbles attributed
to episodic SMBH-driven outbursts~\cite{Pillepich2021}. A relevant reference is the magneto-hydrodynamic bubble-inflation
modelling of Yang, Ruszkowski \& Zweibel~\cite{Yang2022}, a proof
of concept that the giant Fermi and eROSITA bubbles can plausibly
arise from past \SgrA{} jet activity; we adopt their
$\sim 2.6\,\mathrm{Myr}$-post-shutoff scenario as our reference. Their
model, however, does not incorporate the significant
misalignment of the \SgrA{} accretion flow, and therefore of
any associated jet, relative to the Galactic rotation axis.
Recent EHT polarimetric observations favour an inner-flow
orientation substantially tilted from the Galactic
axis~\citep{EHT2024VIII}; sub-parsec stellar disks have
angular-momentum vectors inclined
 relative to the Galactic rotation axis (for a review see
e.~g.~\citep{Genzel2010}); and galaxy-formation simulations
resolving the kpc-to-sub-pc accretion cascade find sub-pc
disks tilted by $0^{\circ}$--$60^{\circ}$ (mean
$\sim 35^{\circ}$) relative to the kpc-scale gaseous
disk~\citep{AnglesAlcazar2021}. Past jets of \SgrA{} are
therefore likely to be significantly misaligned with respect
to the Galactic rotation axis.

Independent hydrodynamical simulations~\cite{Sarkar2023}
explicitly tested a static tilted-jet geometry, motivated by
these same works, and found that the observed axisymmetry and
north--south hemisymmetry of the Fermi/eROSITA bubbles can be
reproduced either by a short ($\lesssim 6\,\mathrm{kyr}$)
super-Eddington outburst ($\gtrsim 5 \times
10^{44}\,\mathrm{erg\,s^{-1}}$) --- disfavoured by the
observed O\,\textsc{viii}/O\,\textsc{vii} line ratio --- or
by a quasi-steady, low-luminosity ($\approx
10^{40.5}$--$10^{41}\,\mathrm{erg\,s^{-1}}$) source --- either
a magnetically dominated jet or accretion wind from \SgrA{},
or a wind driven by supernovae or tidal disruption events in
the Galactic Center --- with
the
magnetically-dominated jet being the regime adopted in our
model.

A static tilted jet, however, naturally introduces tension
with the near-spherical morphology of the GCE: confined to a
single off-axis cone, its hadronic CR injection would
preferentially seed an anisotropic gamma-ray distribution
relative to the observed residual. Resolving this tension
motivates adding precession to the established tilted-jet
picture --- and, fortunately, precession is not an ad-hoc
ingredient but a natural consequence of accreting onto a
spinning black hole (BH) through a misaligned disc:
Lense--Thirring frame-dragging torques the inner disc and
sweeps the jet direction azimuthally around the spin axis on
a timescale set by the disc geometry and the black-hole
spin~\cite{Fragile2007,Liska2018}.

In this paper we therefore propose that a {precessing}
paraboloidal Blandford--Znajek (BZ) jet~\cite{BZ1977} from \SgrA{} during its past active phase drove the bipolar
expansion of the Fermi/eROSITA bubbles --- contributing,
alongside wider-angle outflows, to their observed extent ---
and injected hadronic cosmic-ray (CR) protons into the inner
Galaxy, producing GCE-relevant gamma-ray emission via
inelastic $pp$ collisions~\cite{Crocker2011,Guo2012FB}. This
generalises the static tilted-jet
picture~\cite{Yang2022,Sarkar2023} by adding the precession
that the misaligned-disc geometry naturally produces:
Lense--Thirring precession of the disc broadens the otherwise
narrow off-axis CR injection into a broadened,
azimuthally-symmetric angular distribution that, after
diffusive transport, is compatible with the observed
near-spherical GCE morphology --- while maintaining the
observed large-scale bubble symmetry, because the precession
period ($T_{\rm prec} \lesssim 1.5\,\Myr$) we consider is
shorter than the bubble inflation timescale ($\tact \simeq
7.5\,\Myr$), so the bubbles inflate along the
precession-averaged jet axis rather than the instantaneous
one. Our results depend on this condition $T_{\rm prec} \ll
\tact$ but not on the specific value of $T_{\rm prec}$. The
transition from this banded angular distribution to the
near-spherical GCE morphology is completed by cosmic-ray
diffusion through the central molecular zone (CMZ) and
Galactic bulge, which we model numerically by convolving the
analytic precession-averaged BZ injection with a two-zone
diffusion Green function using diffusion coefficients
consistent with values adopted in inner-Galaxy hadronic
propagation studies. The CMZ acts as a slow-diffusion inner
zone that partially confines CRs near the source, while the
surrounding bulge provides the outer transport zone over
which the cloud spreads radially on a length scale
$\lambda_{\rm bulge} \simeq 1.7\,\kpc$ 
at the Galactocentric
distance $d_\odot\simeq8.2\,\kpc$,
comparable to the observed angular extent of the GCE. The model thereby
connects the axis tilt, the
jet's contribution to the Fermi/eROSITA
inflation, and the GCE residual
within a single self-consistent picture of
\SgrA{}'s recent activity. This precessing-jet treatment of
the CR injection appears not to have been previously combined
with the bubble inflation and GCE morphology in a single
framework.

\begin{figure}[!htbp]
\centering
\includegraphics[width=12cm]{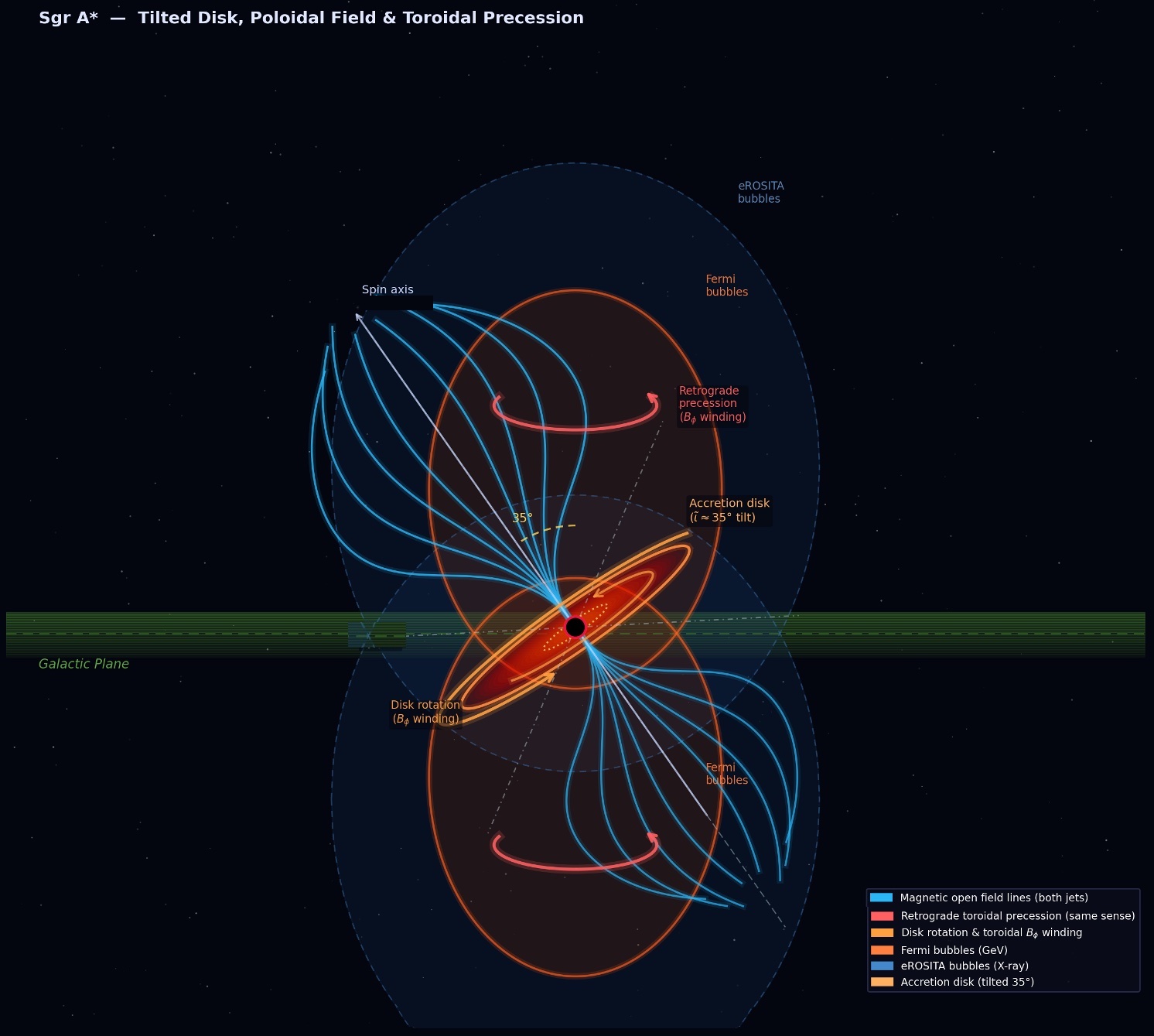}
\caption{Schematic of the proposed model. The accretion disc
(orange ellipse), tilted from the Galactic
rotation axis (perpendicular to the green band representing
the Galactic plane), launches a paraboloidal BZ jet within a
critical cone of half-angle $\theta_{c} \simeq 20^{\circ}$
from the spin axis (dashed grey lines). Open poloidal field
lines (not to scale in cyan) thread the polar caps.
Lense--Thirring precession (red arrows) sweeps the spin axis
azimuthally over the active phase $\tact $,
broadening the cosmic-ray injection into an
azimuthally-symmetric angular distribution that is
subsequently isotropized by diffusion through the CMZ and
Galactic bulge. The Fermi (orange-red lobes) and eROSITA
(blue dashed ellipses) bubbles are interpreted as fossil
evidence of this past outburst,
the jet driving their bipolar expansion
alongside wider-angle outflows that supply the remaining
extent, with their large-scale symmetry preserved by
precession averaging on timescales short compared to the
bubble inflation. The dashed blue line indicates the spin
axis ($\astar \simeq 0.9$); the gold arc marks the
$35^{\circ}$ tilt angle.}
\label{fig:tilted_disk}
\end{figure}

The BZ mechanism, central to our proposal, sits within a
broader theoretical landscape of energy-extraction processes
from rotating black holes, sometimes referred to more broadly
as superradiance. The original Penrose
process~\cite{Penrose1969,PenroseFloyd1971} demonstrated that
the rotational energy of a Kerr black hole can in principle
be tapped through orbits within the ergosphere;
astrophysically relevant variants include superradiant
amplification of fields~\cite{ZelDovich1971,StarobinskyChurilov1973,PressTeukolsky1972,Brito2015review},
and electromagnetic energy extraction by the BZ
mechanism~\cite{BZ1977,Wald1974,Beskin2010,Gralla2016jets,Lupsasca2014}.
Numerical simulations of magnetically arrested accretion
confirm that BZ extraction can convert a substantial fraction
of the black-hole rotational energy into a Poynting-dominated
jet~\cite{Tchekhovskoy2011,Narayan2012MAD,McKinney2012MCAFs,Liska2020,Ripperda2022},
with conversion efficiency $\eta_{\BZ}$ approaching unity for
$a_{*} \gtrsim 0.9$ and saturated horizon flux; the resulting
jet power and total energy budget are derived in
Sec.~\ref{ssec:BH}.

The paper is structured as follows.
Section~\ref{sec:model} describes the physical model.
Section~\ref{sec:derivation} presents the derivation
of the hadronic gamma-ray surface brightness, introducing the
isotropic CR-cloud approximation, solving the two-zone
diffusion problem numerically for the cosmic-ray profile, and
integrating along the line of sight to obtain the
energy-resolved spectrum and brightness.
Section~\ref{sec:chi2} compares the results against current
GCE measurements and quantifies the spin-dependent hadronic floor. Section~\ref{sec:triaxial} develops the
triaxial-bar model to the gas density and the resulting
peaked longitudinal asymmetry as a categorical observational
discriminator. Section~\ref{sec:conclusions} summarizes our
results, and discusses consistency with present-day
observations of \SgrA{} and the inner Galaxy.
Appendix~\ref{app:MagneticField} collects the analytical
structure of the BZ magnetic field.
Appendix~\ref{app:multipole} presents the formal multipole
analysis of the diffusion equation that justifies the
isotropic-cloud approximation used in the main text.
Appendix~\ref{app:greenbox} gives the closed-form
quasi-stationary two-zone solution used to validate the
numerical cosmic-ray profile.

\section{The Physical Model}
\label{sec:model}

We model \SgrA{} as a magnetically-dominated, sub-Eddington
source that {contributes to both} the
Fermi/eROSITA bubbles and the GCE through three coupled
processes: (i)~a precessing paraboloidal BZ jet depositing
hadronic CRs within a polar cone; (ii)~Lense--Thirring
precession of a tilted accretion disc sweeping that cone over
the full azimuth; (iii)~two-zone CR diffusion redistributing
the injection into a morphology matching the observed GCE.
Each ingredient is independently constrained by observation;
the combination determines the jet's
contribution to the bubble geometry, the GCE spatial
profile, and an irreducible hadronic floor.

\begin{figure}[!htbp]
\centering
\includegraphics[width=12cm]{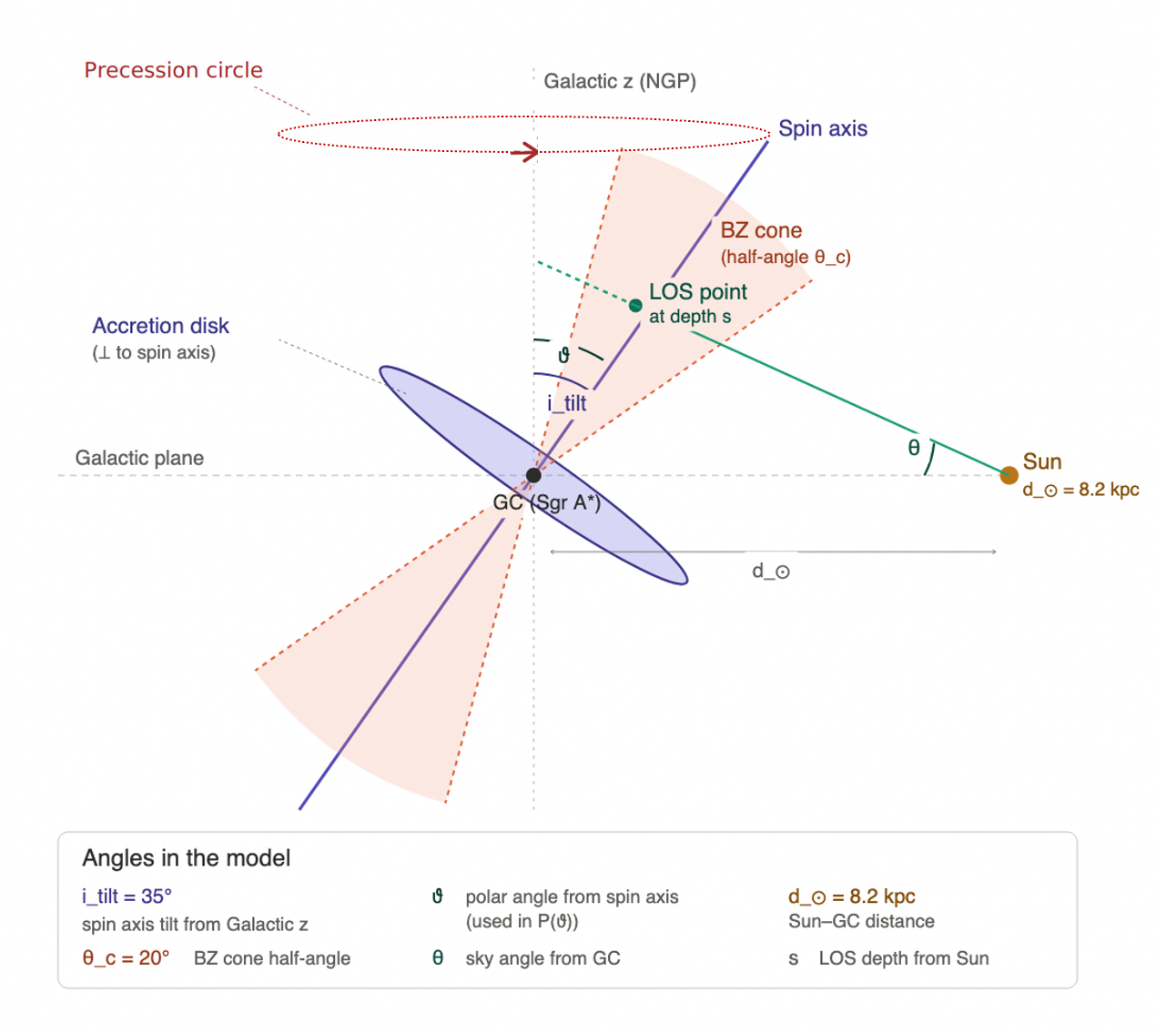}
\caption{Horizon-scale geometry of the precessing BZ jet from
\SgrA{}. The spin axis (purple, dashed) is tilted by
$i_{\rm tilt} \simeq 35^{\circ}$ from the Galactic
$\hat{z}$-axis, consistent with \citep{Genzel2010}. The
accretion disc (purple ellipse) lies perpendicular to the
spin axis; the BZ critical cone (orange wedge) is launched
with half-angle $\theta_{c} \simeq 20^{\circ}$ around the
instantaneous spin axis. The Lense--Thirring precession
circle (red dashed ellipse) traces the spin-axis tip around
the Galactic $\hat{z}$-axis. The polar angle $\vartheta$
(from Galactic $\hat{z}$) and sky angle $\theta$ (from the
Sun--GC line of sight) entering Eq.~(\ref{eq:Pplus}) and the
surface-brightness integral of Sec.~\ref{sec:derivation} are
indicated.}
\label{fig:geometry}
\end{figure}

The black-hole and jet-power parameters are summarized in
Sec.~\ref{ssec:BH}; the horizon-scale magnetic field is
derived in Sec.~\ref{ssec:Bfield}, with the full vector
potential given in Appendix~\ref{app:MagneticField}; the jet
geometry, tilt and Lense--Thirring precession are treated in
Sec.~\ref{ssec:precession}; the precession-averaged CR
injection fraction is derived in Sec.~\ref{ssec:injection};
the jet-only contribution to the bubble
extent in Sec.~\ref{ssec:jet_single_source}; and the
resulting hadronic gamma-ray budget in
Sec.~\ref{ssec:budget}. The geometry of the model ---
including the spin-axis tilt $i_{\tilt}$, the BZ cone
half-angle $\theta_{c}$, the precession circle, the accretion
disc orientation, and the line-of-sight angles $\theta$ and
$\vartheta$ entering the surface-brightness integral of
Sec.~\ref{sec:derivation} --- is represented in
Fig.~\ref{fig:geometry}, to which we refer throughout.

\subsection{Black-hole and jet parameters}
\label{ssec:BH}

\SgrA{} is modelled as a Kerr black hole of mass $\MBH =
4.15\times10^{6}\,\Msun$, a representative value within the
range of recent dynamical mass determinations
\citep{Ghez2008, Do2019, GRAVITY2022} and dimensionless
spin $\astar = 0.9$, consistent with the outflow-method
analysis of~\cite{Daly2024} ($\astar = 0.90 \pm 0.06$) and
the EHT 2024 polarimetric preferred model~\cite{EHT2024VIII}
($\astar = 0.94$, magnetically arrested disc (MAD),
$i = 150^{\circ}$). The accretion flow is geometrically thick
and radiatively inefficient --- an
ADAF~\cite{Yuan2003,Yuan2014ARAA} --- appropriate for the
inferred sub-Eddington accretion rate. Such flows efficiently
advect large-scale poloidal magnetic flux
inward~\cite{Tchekhovskoy2011,Narayan2012MAD}, until the
accumulated flux at the horizon becomes strong enough to
impede further accretion. We assume the MAD state is
sustained throughout the active phase, during which a
BZ-powered jet is continuously launched. The gravitational
and event-horizon radii are
\begin{equation}
  r_{g} = \frac{G\MBH}{c^{2}} = 6.13\times10^{11}\,
                                  \mathrm{cm},
  \quad
  r_{+} = r_{g}\!\left(1+\sqrt{1-\astar^{2}}\right)
        = 1.44\,r_{g},
  \label{eq:Rg}
\end{equation}
giving horizon angular velocity
\begin{equation}
  \Omega_{H} = \frac{\astar c}{2 r_{+}}
             = 1.54\times10^{-2}\,\mathrm{rad\,s^{-1}}.
  \label{eq:OmegaH}
\end{equation}

The Eddington luminosity is
\begin{equation}
  L_{\Edd} = \frac{4\pi G \MBH m_p c}{\sigma_T}
          = 1.26\times10^{38}\,\frac{\MBH}{\Msun}\,\ergs
          = 5.23\times10^{44}\,\ergs,
  \label{eq:LEdd}
\end{equation}
where $m_p$ is the proton mass and $\sigma_T$ is the Thomson
cross-section. The corresponding Eddington accretion rate is
\begin{equation}
  \Mdot_{\Edd} \;\equiv\; \frac{L_{\Edd}}{\eta_{\rm rad}\,c^{2}}
              \;\simeq\; 0.092\,\Msun\,\mathrm{yr}^{-1},
  \label{eq:MdotEdd}
\end{equation}
assuming a fiducial radiative efficiency
$\eta_{\rm rad} = 0.1$ appropriate for a standard disc
accretion regime. During the active phase the
Eddington-scaled accretion rate is fixed at
$\mdot \equiv \Mdot/\Mdot_{\Edd} = 10^{-5}$ (with
corresponding mass accretion rate $\Mdot \simeq
9.2\times10^{-7}\,\Msun\,\mathrm{yr}^{-1}$), a $\sim 10^{3}$
enhancement over the present-day quiescent value
$\mdot_{\rm today} \sim 10^{-8}$ ~\citep{EHT2024VIII}. These
values are comparable in Eddington units to M87's
present-day horizon-scale rate, $\dot{m}_{\rm M87} \approx
(2$--$14)\times 10^{-6}$~\citep{EHT2021VIII} --- both deep in
the ADAF/RIAF regime.

In contrast to the accretion rate, the BH spin is essentially
frozen over the timescales relevant to this work. The
fractional mass --- and hence spin --- change over the active
phase is
\begin{equation}
\frac{\Delta \astar}{\astar}
\;\lesssim\;
\frac{\Delta \MBH}{\MBH}
\;=\;
\frac{\Mdot \cdot \tact}{\MBH}
\;\sim\; 2\times10^{-6},
\end{equation}
entirely negligible. The spin is therefore treated as
constant between the active phase and today, making it the
most robustly time-fixed physical parameter in the model.

The BZ jet power is given by the standard
relation~\cite{BZ1977,Tchekhovskoy2011}
\begin{equation}
  P_{\BZ} = \eta_{\BZ}\,\Mdot c^{2}
  \label{eq:PBZ}
\end{equation}
where $\eta_{\BZ}$ is the dimensionless BZ efficiency. The BZ
efficiency
\begin{equation}
\eta_{\BZ}
=\frac{\kappa }{4\pi} \phi^2 \, \omega_H^2
  \left(1 + 1.38\,\omega_H^2 - 9.2\,\omega_H^4\right)
  \label{eq:eta}
\end{equation}
with dimensionless horizon flux $\phi$ and angular velocity
defined as $\omega_{H}= \Omega_{H} \,r_{g} /c$ follows the
MAD scaling ~\cite{Tchekhovskoy20102,Tchekhovskoy2011}. For
a high-spin ($\astar = 0.9$) black hole at MAD saturation,
GRMHD simulations give $\eta_{\BZ}\simeq
1$~\cite{Tchekhovskoy2011,Narayan2012MAD}. With the fiducial
$\mdot = 10^{-5}$, this yields
\begin{equation}
  P_{\BZ} \simeq 5.23\times 10^{40}\,\ergs,
  \label{eq:valuePBZ}
\end{equation}
which corresponds to the total bipolar power.\footnote{At a
more conservative $\astar = 0.5$, $\eta_{\BZ} \simeq 0.18$
and the floor amplitude drops by factor $\sim 5$.} The total
injected jet energy over the fiducial active duration
$\tact = 7.5\,\Myr$ is
\begin{equation}
  E_{\mathrm{jet}} = P_{\BZ}\,\tact
                  \approx 1.2\times 10^{55}\,\erg,
  \label{eq:Ejet}
\end{equation}
within the $\sim 10^{55}$--$10^{57}\,\erg$ range of published
Fermi/eROSITA bubble energy
estimates~\cite{Su2010Fermi,Predehl2020,Yang2022}.
Complementary energy sources proposed in the literature
include accretion-disc winds~\cite{MouYuan2014} and repeated
TDE-driven outflows~\cite{Ko2020}, whose
combined contribution to the full bubble extent is discussed
in the context of the jet-only estimate of
Sec.~\ref{ssec:jet_single_source}.

\subsection{Magnetic field profile and horizon field}
\label{ssec:Bfield}

The horizon magnetic field strength is set by the
steady-state MAD saturation condition. In the MAD
state~\cite{Tchekhovskoy2011,Narayan2012MAD}, large-scale
poloidal flux accumulates at the BH horizon until magnetic
stresses periodically arrest the accretion inflow. The
saturation level is characterized by the dimensionless
horizon flux
\begin{equation}
  \phi
  \;\equiv\;
  \frac{\Phi_{\mathrm{BH}}}{\sqrt{\dot M c\, r_{g}^{2}}}
  \;\simeq\; 50,
  \label{eq:phiMAD}
\end{equation}
where $\Phi_{\mathrm{BH}} = \int_{H}\!\bm B\!\cdot\!d\bm A$
is the poloidal flux threading one hemisphere of the
horizon~\cite{Tchekhovskoy2011}. The horizon flux
$\Phi_{\mathrm{BH}}$ is determined by the BZ jet power
through the BZ relation
\begin{equation}
  P_{\BZ}
  \;=\;
  \frac{\kappa}{4\pi c}\,{\Omega_{H}^{2}\,\Phi_{\mathrm{BH}}^{2}}\,,
  \label{eq:PBZ1}
\end{equation}
with BZ geometric coefficient $\kappa \simeq 0.045$ (see
App.~\ref{app:MagneticField} for details). The two equivalent
forms~\eqref{eq:PBZ} and~\eqref{eq:PBZ1} for $P_{\BZ}$ are
related at MAD saturation through \eqref{eq:eta}. Inverting
Eq.~(\ref{eq:PBZ1}) for the chosen jet power $P_{\BZ}$ fixes
the active-phase horizon flux, and hence the field
\begin{equation}
  B_{H}^{(0)}
  \;\equiv\;
  \frac{\Phi_{\mathrm{BH}}}{\pi r_{+}^{2}}
  \;=\;
  \frac{1}{\pi r_{+}^{2}}
  \sqrt{\frac{4\pi c\,P_{\BZ}}{\kappa\,\Omega_{H}^{2}}}
  \;\simeq\; 2 \times 10^{4}\,\mathrm{G},
  \label{eq:BH0}
\end{equation}

The active-phase $B_{H}^{(0)}$ is much larger than the
present-day quiescent value~\cite{Gravity,EHT2024VIII},
reflecting the transition from MAD saturation to sub-MAD
accretion after jet shutoff. After jet shutoff at
$t = \tact$, the horizon magnetic flux is no longer sustained
by accretion. The system transitions from a MAD state to a
progressively weaker, sub-MAD configuration as magnetic flux
is advected outward and dissipated through reconnection and
turbulent diffusion. The post-shutoff value of the horizon
magnetic field is not used in the downstream GCE calculation.

\paragraph{Jet field beyond the collimation break.}
The jet structure we consider consists of two regimes. Close
to the black hole the jet follows a paraboloidal streamline,
$r_{\perp}(s) \propto s^{k}$ with $k \simeq
0.58$ and $s$ the distance measured along the jet axis ~\cite{AsadaNakamura2012}, collimated by external gas
bound to the BH's gravitational
potential~\cite{Komissarov2009,Tchekhovskoy2008}. Further
downstream, beyond the collimation break, the jet transitions
into a freely-expanding conical shape. In M87 this transition
is observed at $s_{\rm break} \simeq 2.5 \times
10^{5}\,R_{S}$ ($\simeq 150\,\mathrm{pc}$, inside the M87
Bondi radius)~\cite{AsadaNakamura2012,Nakamura2018}; the same
mechanism is expected to operate in the \SgrA{} MAD jet
during the active phase, placing the corresponding transition
near the \SgrA{} Bondi radius $r_{B} \sim
0.04\,\mathrm{pc}$. Beyond $s_{\rm break}$, the jet
propagates as a free conical flow with cylindrical radius
$r_{\perp}(s) = s\,\tan\theta_{\rm open}$ measured from the
instantaneous jet axis, which precesses about Galactic
$\hat{z}$ on the period $T_{\mathrm{prec}}$
(Sec.~\ref{ssec:precession}). The asymptotic Poynting flux is
approximately equipartitioned between the toroidal magnetic
and kinetic components~\cite{Tchekhovskoy2011,Komissarov2007};
equating the Poynting flux of a single jet carried at
$\sim c$ through cross-sectional area $\pi r_{\perp}^{2}$ to
$P_{\BZ}/2$ gives
\begin{equation}
  B_{\rm jet}(r_{\perp})
  \;=\;
  \left(\frac{4\,P_{\BZ}}{c\,r_{\perp}^{2}}\right)^{1/2},
  \label{eq:Bjet}
\end{equation}
where $P_{\BZ}$ is the total bipolar BZ power
(Sec.~\ref{ssec:BH}). For 
\eqref{eq:valuePBZ}
this yields $B_{\rm jet} \simeq
100\,\mu\mathrm{G}$ at $s = 0.1\,\kpc$ ($r_{\perp} \simeq
9\,\mathrm{pc}$), $\simeq 10\,\mu\mathrm{G}$ at $s =
1.0\,\kpc$ ($r_{\perp} \simeq 90\,\mathrm{pc}$), and $\simeq
1.9\,\mu\mathrm{G}$ at the bubble wall ($s \simeq 5\,\kpc$,
$r_{\perp} \simeq 0.44\,\kpc$). These are instantaneous
on-axis values along the precessing jet axis.
At the forward shock the ambient field is
additionally amplified by strong-shock compression, so the total
post-shock wall field used in the Hillas estimate of
Sec.~\ref{ssec:budget} is $B_{\wall} \simeq
4\,\mu\mathrm{G}$ consistent with the few-microgauss fields inferred for
the Fermi bubbles.

\subsection{Jet geometry: tilt and Lense--Thirring
            precession}
\label{ssec:precession}

We model a permanently tilted, precessing disc--jet system
maintained at fixed misalignment with respect to the BH spin
axis. We adopt $i_{\tilt} = 35^{\circ}$ as fiducial. A
misaligned accretion disc around a spinning BH experiences
Lense--Thirring frame dragging, which torques tilted orbits
at every radius~\cite{LenseThirring1918}. The
disc response depends on its aspect ratio $H/r$ -- disc scale height over radius. For a thin disc $H/r \ll \alpha$, where $\alpha$ is the
Shakura--Sunyaev viscosity parameter, the dominant outcome is
Bardeen--Petterson alignment of the inner disc with the BH
spin equator on a viscous
timescale~\cite{Bardeen1975}; for a
thick disc $H/r \gtrsim \alpha$, warp information
propagates as bending waves rather than diffusively, and the
inner disc--jet system instead undergoes coherent rigid-body
precession about the BH spin
axis~\cite{Fragile2007,Liska2018}. The \SgrA{} RIAF, with
$H/r \sim 0.3$ and $\alpha \sim 0.1$~\cite{Yuan2014ARAA},
sits well inside the thick-disc regime and the Bardeen--Petterson
alignment is suppressed. This thick-disc precession mechanism has observational support in
M87, where Cui et al.~\cite{Cui2023M87} report an $\sim 11$-yr
periodicity in the jet position angle, interpreted as
Lense--Thirring precession of a misaligned accretion disc around a
spinning black hole. M87's accretion flow, like that of \SgrA{}, is
a hot, geometrically-thick RIAF, so this provides a precedent.

\subsection{Precession-averaged CR injection fraction}
\label{ssec:injection}

For a precessing time $T_{\rm prec} \ll
\tact$ the time-dependent BZ cone is replaced by
its azimuthal average. The jet deposits CRs within half-angle
$\theta_{c} = 20^{\circ}$ of its instantaneous axis, the
horizon-scale BZ critical cone set by the angular structure
of the paraboloidal poloidal field. For a direction at Galactic
polar angle $\vartheta$ measured from the North Galactic
Pole, the fraction of the precession period during which the
upper BZ cone illuminates that direction is
\begin{equation}
  P^{+}(\vartheta) = \frac{1}{\pi}\arccos\!\left(
      \frac{\cos\theta_{c} - \cos\vartheta\cos i_{\tilt}}
           {\sin\vartheta\,\sin i_{\tilt}}\right),
  \label{eq:Pplus}
\end{equation}
for $|\cos\theta_{c} - \cos\vartheta\cos i_{\tilt}| \le
\sin\vartheta\sin i_{\tilt}$; otherwise $P^{+} = 0$ or $1$.
The bipolar injection fraction, including the southern
counter-jet $P^{-}(\vartheta) = P^{+}(\pi-\vartheta)$, is
\begin{equation}
  P(\vartheta) = P^{+}(\vartheta) + P^{-}(\vartheta)
                 - P^{+}(\vartheta)\,P^{-}(\vartheta).
  \label{eq:Ptot}
\end{equation}
For fiducial $(i_{\tilt}, \theta_{c}) = (35^{\circ},
20^{\circ})$, $P^{+}$ has support on
$\vartheta \in [15^{\circ}, 55^{\circ}]$ and $P^{-}$ on the
disjoint conjugate range $[125^{\circ}, 165^{\circ}]$, so the
cross term vanishes identically and $P(\vartheta) =
P^{+}(\vartheta) + P^{-}(\vartheta)$. The resulting annular
injection profile can reproduce the morphological scale of
the observed GCE~\cite{DiMauro2021,Calore2015}.

\subsection{Jet contribution of the Fermi/eROSITA bubbles}
\label{ssec:jet_single_source}
Of the bipolar power $P_{\BZ}$, the dominant share is advected
along the jet channel at relativistic speed and thermalized at
the working surface, inflating the lobes; a fraction
$\xi_{\CR} \simeq 0.05$ is converted into non-thermal protons
within the inner $\sim 100\,\mathrm{pc}$
(Sec.~\ref{ssec:budget}). Here we focus on the larger-scale
structures driven by the dominant mechanical share, deferring
the transport of the CR components to
Sec.~\ref{sec:derivation}.

The jet inflates the lobe in two phases, joined continuously
at $t=\tact$: during the active phase ($t<\tact$) energy
accumulates as $E(t)=P_{\BZ}^{\rm lobe}\,t$, with
$P_{\BZ}^{\rm lobe}\equiv\tfrac12(1-\xi_{\CR})\,P_{\BZ}$ the
mechanical power per lobe, and the shock is continuously
driven; after shutoff ($t>\tact$) the energy is fixed at
$E_{\rm jet}^{\rm lobe}=P_{\BZ}^{\rm lobe}\,\tact
\simeq 5.9\times10^{54}\,\erg$ and the shock coasts. In a stratified medium $\rho(r)=\rho_{0}(r/r_{0})^{-k}$
the self-similar forward-shock radius obeys the single law
\begin{equation}
  R(t)=\xi\,
  \left(\frac{\mathcal{C}\,E_{\rm eff}(t)\,t^{2}}
             {\rho_{0}\,r_{0}^{-k}}\right)^{\!1/(5-k)},
  \qquad
  E_{\rm eff}(t)=
  \begin{cases}
  P_{\BZ}^{\rm lobe}\,t, & t<\tact,\\
  E_{\rm jet}^{\rm lobe}, & t>\tact,
  \end{cases}
  \label{eq:Rgeneral}
\end{equation}
where $k$ is the density power-law slope ($\rho\propto r^{-k}$),
$\mathcal{C}$ the directional concentration factor, and $\xi$ the
$\mathcal{O}(1)$ Sedov constant for that direction.

This concentration factor arises because the precessing jet does
not inject isotropically: it sweeps a bipolar band of total solid
angle $\Omega_{\rm tot}=4\pi[\cos(i_{\tilt}-\theta_c)
-\cos(i_{\tilt}+\theta_c)]$, i.e.\ a sky fraction
$f_\Omega\equiv\Omega_{\rm tot}/4\pi\simeq0.39$ for the fiducial
tilt $i_{\tilt}=35^\circ$ and precession-cone half-angle
$\theta_c\simeq20^\circ$. Since $P_{\BZ}$ is the power averaged over
the full $4\pi$, confining it to this band raises the energy per
steradian along the axis by
\begin{equation}
  \mathcal{C}_{\parallel}\equiv\frac{4\pi}{\Omega_{\rm tot}}
  =\frac{1}{f_\Omega}\simeq2.55,
  \qquad
  \mathcal{C}_{\perp}\equiv\mathcal{C}_{\parallel}^{-1/2}\simeq0.63,
\end{equation}
with the closure $\mathcal{C}_{\parallel}\mathcal{C}_{\perp}^{2}=1$
conserving the total energy (axial gain balanced by lateral
reduction); an isotropic source has
$\mathcal{C}_{\parallel}=\mathcal{C}_{\perp}=1$.

The vertical and lateral extents follow by setting
$k_{\parallel}=3/2$ (the stratified halo, $\rho\propto r^{-3/2}$,
i.e.\ a  hot-gas
model~\cite{MillerBregman2015,Zhang2020})and $k_{\perp}=0$
(uniform ambient density across the bubble waist), with density
$\rho_{0}=\mu m_{\rm H} n_{0}\simeq
3\times10^{-27}\,\mathrm{g\,cm^{-3}}$ at $r_{0}=1\,\kpc$
($\mu\simeq0.6$, $n_{0}=3\times10^{-3}\,\mathrm{cm^{-3}}$). This
yields
\begin{equation}
  R_{\parallel}^{\rm jet}(t) =
  \begin{cases}
  \xi_{\parallel}^{\rm act}\!
    \left(\dfrac{\mathcal{C}_{\parallel}P_{\BZ}^{\rm lobe}t^{3}}
                {\rho_{0}r_{0}^{3/2}}\right)^{\!2/7}
    \!\!\propto t^{6/7}, & t<\tact,\\[2.4ex]
  \xi_{\parallel}^{\rm off}\!
    \left(\dfrac{\mathcal{C}_{\parallel}E_{\rm jet}^{\rm lobe}t^{2}}
                {\rho_{0}r_{0}^{3/2}}\right)^{\!2/7}
    \!\!\propto t^{4/7}, & t>\tact,
  \end{cases}
  \label{eq:Rpar}
\end{equation}
\begin{equation}
  R_{\perp}^{\rm jet}(t) =
  \begin{cases}
  \xi_{\perp}^{\rm act}\!
    \left(\dfrac{\mathcal{C}_{\perp}P_{\BZ}^{\rm lobe}t^{3}}
                {\rho_{0}}\right)^{\!1/5}
    \!\!\propto t^{3/5}, & t<\tact,\\[2.4ex]
  \xi_{\perp}^{\rm off}\!
    \left(\dfrac{\mathcal{C}_{\perp}E_{\rm jet}^{\rm lobe}t^{2}}
                {\rho_{0}}\right)^{\!1/5}
    \!\!\propto t^{2/5}, & t>\tact .
  \end{cases}
  \label{eq:Rperp}
\end{equation}
The $\xi$ are the $\mathcal{O}(1)$ constants of the
self-similar solution for ~\cite{Ostriker1988ST}; we fix their vertical-to-lateral ratio
by the observed eROSITA forward-shock aspect ratio ($\simeq2{:}1$),
giving $\xi_{\parallel}^{\rm off}\simeq0.81$ and
$\xi_{\perp}^{\rm off}\simeq1.02$ for the coasting (impulsive)
branch, with the constants
$\xi_{\parallel}^{\rm act}\simeq0.83$,
$\xi_{\perp}^{\rm act}\simeq1.05$ a few percent larger, as expected
for a continuously-driven shock. Since
$E_{\rm jet}^{\rm lobe}=P_{\BZ}^{\rm lobe}\tact$, the driven and
coasting branches share the same argument at $t=\tact$, differing only through their Sedov constants. These agree to $\sim3\%$, so
the two solutions match at shutoff to that level and the radius is
continuous across the transition. The steeper vertical falloff ($k_{\parallel}=3/2$,
exponent $2/7$) and the injection anisotropy
($\mathcal{C}_{\parallel}/\mathcal{C}_{\perp}\simeq4$) contribute
comparably to the elongation, each enhancing the axial-to-lateral
ratio by $\sim40\%$ relative to an isotropic source in a uniform
medium.

A {single} active duration $\tact=7.5\,\Myr$ followed by a
coasting phase $T_{\rm off}=2.6\,\Myr$ (so $T_{\rm now}=10.1\,\Myr$),
with \eqref{eq:valuePBZ} and the halo above yields the forward-shock radii
\begin{equation}
  R_{\parallel,\rm eROSITA}^{\rm jet}\simeq9.9\,\kpc,
  \qquad
  R_{\perp,\rm eROSITA}^{\rm jet}\simeq3.5\,\kpc
  \quad(\text{aspect } \simeq2),
\end{equation}
The $\gamma$-ray Fermi emission
instead traces the contact discontinuity (CD) at
$ R_{\rm Fermi}^{\rm jet}=\eta\, R_{\rm eROSITA}^{\rm jet}$, lying behind the shock by the
shocked-shell thickness. This thickness is
direction-dependent --- the strong vertical shock leaves a thin
shell (CD near the shock), the weaker lateral shock a thick one ---
so we take
\begin{equation}
  \eta_{\parallel}\equiv
  \frac{R_{\parallel}^{\rm CD}}{R_{\parallel}^{\rm sh}}\simeq0.64,
  \qquad
  \eta_{\perp}\equiv
  \frac{R_{\perp}^{\rm CD}}{R_{\perp}^{\rm sh}}\simeq0.43,
\end{equation}
as expected for an elongated jet-driven lobe where the vertical working surface advances faster than spherical self-similarity assumes. Their ratio
$\eta_{\parallel}/\eta_{\perp}\simeq1.5$ converts the shock aspect
ratio $\simeq2$ into the CD aspect ratio $\simeq3$, giving
\begin{equation}
  R_{\parallel,\rm Fermi}^{\rm jet}\simeq4.4\,\kpc,
  \qquad
  R_{\perp,\rm Fermi}^{\rm jet}\simeq1.5\,\kpc
  \quad(\text{aspect } \simeq3).
\end{equation}
The same jet lifetime thus reproduces {both} bubble systems: its forward shock matches the eROSITA extent and
$\simeq2{:}1$ aspect ratio, while the CD it bounds matches the Fermi
extent and $\simeq3{:}1$ aspect ratio. The jet alone reaches only part of the full vertical extent; the
remainder could plausibly arise from additional sources (e.g.\
wider-angle outflows or nuclear winds; Sec.~\ref{ssec:BH}).

\subsection{Hadronic CR injection and the GCE budget}
\label{ssec:budget}
In addition to the large-scale lobes, the precessing BZ jet
injects relativistic CR protons at the GC over the active phase
$\tact$, with total energy
\begin{equation}
  E_{\CR} = \xi_{\CR}\,P_{\BZ}\,\tact
  \simeq 6.2\times 10^{53}\,\erg
  \qquad (\xi_{\CR} = 0.05),
  \label{eq:ECR}
\end{equation}
a conservative choice within the diffusive-shock-acceleration
range $\xi_{\CR}\in[0.01,0.3]$~\cite{Caprioli2014,Caprioli2012,
Berezhko1997}; the hadronic floor derived below scales linearly
with $\xi_{\CR}$. We model only the hadronic channel: a leptonic
component is present but, cooling within the bubble interior,
contributes negligibly to the GCE region
specifically.\footnote{Leptons ($e^\pm$ from BZ pair production
or shock acceleration) cool in $\sim1$--$10\,\Myr$ via inverse
Compton and synchrotron ($B\sim5\,\mu$G), tracing the bubble
interior rather than the diffusively-spread protons.}

These protons produce gamma rays through
$pp\to\pi^0\to\gamma\gamma$ on the ambient gas. The system is
deeply {thin-target}: at the bulge density
$n_{\rm eff}\sim0.03\,\mathrm{cm^{-3}}$ the $pp$ loss time
exceeds $\tact$ by $\sim10^2$, so only a small fraction of the
proton energy is radiated. The corresponding energy-budget
estimate,
$L_\gamma\simeq\tfrac13 n_{\rm eff}\sigma_{pp}c\,E_{\CR}$,
gives $L_\gamma/L_{\GCE}\simeq11$--$54\%$ across
$n_{\rm eff}\in[0.01,0.05]\,\mathrm{cm^{-3}}$ (relative to
$L_{\GCE}\simeq2.3\times10^{37}\,\ergs$~\cite{DiMauro2021,
Daylan2016}). This is an {upper bound}; the
morphology-matched result of Sec.~\ref{sec:chi2} tightens it to
a hadronic floor at the $\sim3$--$14\%$ level.

The morphology is set by how the protons spread between
injection and decay. They diffuse through two zones --- a
slow-diffusion CMZ ($r<200\,\mathrm{pc}$,
$D_{\rm CMZ}=10^{27}\,\mathrm{cm^2\,s^{-1}}$) embedded in the
faster Galactic bulge ($D_{\rm bulge}=3\times10^{28}\,
\mathrm{cm^2\,s^{-1}}$) --- a picture consistent with the
$\propto1/r$ CR profile HESS measures across the
CMZ~\cite{HESS2016}, and with values from dedicated
Galactic Center transport models~\cite{GabiciAharonianBlasi2007,
Scherer2023,Gaggero2015}. 
Sec.~\ref{sec:derivation} develops the full transport
calculation that yields the realized floor, constrained in
Sec.~\ref{sec:chi2}.

\section{Derivation of the Hadronic Gamma-Ray Surface Brightness}
\label{sec:derivation}

In this section we derive the gamma-ray surface brightness
produced by hadronic CR proton interactions in the inner
Galaxy. We begin with the volume emissivity
(Sec.~\ref{ssec:emissivity}) and the isotropic CR cloud
approximation (Sec.~\ref{ssec:factorisation}), obtain the
radial CR profile from the two-zone transport calculation
(Sec.~\ref{ssec:Green}) and the energy-resolved proton spectrum
from a leaky-box treatment (Sec.~\ref{ssec:spectrum}), and
project these into the line-of-sight surface brightness
(Sec.~\ref{ssec:Ib}) and the region-averaged GCE spectrum
(Sec.~\ref{ssec:spectrum_recipe}). The numerical implementation
and fiducial parameters are collected in
Sec.~\ref{ssec:numerics}. The precession-averaged angular factor
$P(\vartheta)$ derived in Sec.~\ref{ssec:injection} enters the
formal multipole analysis of Appendix~\ref{app:multipole}; we
show in Sec.~\ref{ssec:factorisation} that diffusive smearing
washes it out at GCE-relevant radii, so the fiducial calculation
 adopts a fully isotropic CR cloud. 

\subsection{Volume emissivity from \texorpdfstring{$pp \to \pi^0 \to \gamma\gamma$}{pp -> pi0 -> gamma gamma}}
\label{ssec:emissivity}

For cosmic-ray protons interacting with diffuse interstellar
gas, the dominant gamma-ray production channel above
$E_{\gamma} \simeq 0.3\,\GeV$ is inelastic proton-proton
scattering producing neutral pions which decay to photon
pairs~\cite{Stecker1971,Dermer1986}:
\begin{equation}
  p + p \;\to\; X + \pi^{0},
  \qquad \pi^{0} \;\to\; \gamma + \gamma,
  \label{eq:pp_pi0}
\end{equation}
where $X$ denotes additional secondary particles. For a CR
proton population with differential number density
$n_{\CR}(E_{p}, \bm{x})$ ($\mathrm{cm^{-3}\,GeV^{-1}}$)
traversing target gas of density $n_{\mathrm{gas}}(\bm{x})$,
the differential photon volume emissivity is~\cite{Kafexhiu2014}
\begin{equation}
  \varepsilon(E_{\gamma}, \bm{x})
  = c\,n_{\mathrm{gas}}(\bm{x})
    \int_{E_{\mathrm{th}}}^{\infty}
    \frac{d\sigma_{pp}(E_{p}, E_{\gamma})}{dE_{\gamma}}\,
    n_{\CR}(E_{p}, \bm{x})\,dE_{p},
  \label{eq:emissivity}
\end{equation}
in $\mathrm{ph\,cm^{-3}\,s^{-1}\,GeV^{-1}}$, where
$E_{\mathrm{th}}$ is the proton kinetic threshold for
$\pi^{0}$ production. We use the Kafexhiu et
al.~\cite{Kafexhiu2014} parametrisation of
$d\sigma_{pp}/dE_{\gamma}$, valid from threshold to PeV and
reproducing accelerator data to better than $20\%$.

Energy losses during the active phase are negligible
($t_{pp} \gg \tact$, Sec.~\ref{ssec:budget}). We adopt a
factorisation ansatz in which the CR spectral shape is treated
as approximately position-independent throughout the bulge.
Because the calculation is normalized by the injected CR
{energy} rather than particle number, and because the
numerical implementation (Sec.~\ref{ssec:numerics}) works
directly with the energy density, we write the factorisation
in energy-normalized form,
\begin{equation}
  n_{\CR}(E_{p}, \bm{x}) =
  {u_{\CR}(\bm{x})} \,\frac{\widetilde{q}(E_{p})}{E_{p}},
  \qquad
  \int E_{p}\,\widetilde{q}(E_{p})\,dE_{p} = 1,
  \label{eq:nCR_factor}
\end{equation}
where $u_{\CR}(\bm{x})$ is the local CR energy density
($\mathrm{erg\,cm^{-3}}$) from Sec.~\ref{ssec:Green} and
$\widetilde{q}(E_{p})$ is the normalized spectral shape
($\mathrm{GeV^{-1}}$) from Sec.~\ref{ssec:spectrum}, fixed so
that the proton population it describes carries unit total
energy. The factorized treatment is standard in inner-Galaxy
hadronic gamma-ray analyses~\cite{Cesarsky1980,Strong2007};
the leading correction from energy-dependent spatial spreading
is expected to remain at the $\sim 30\%$ level.

The total CR energy injected during $\tact$ is fixed by the
BZ jet's power budget, Eq.~(\ref{eq:ECR}),
\begin{equation}
  E_{\CR} \;=\; \xi_{\CR}\,P_{\BZ}\,\tact
        \;=\; \int u_{\CR}(\bm{x})\,d^{3}x
        \;\simeq\; 6.2\times10^{53}\,\erg
  \label{eq:ECR_def}
\end{equation}
at $\xi_{\CR} = 0.05$; this single
scalar carries the entire dependence on black-hole spin, jet
power, and active duration into the gamma-ray calculation,
entering nowhere else. Substituting Eq.~(\ref{eq:nCR_factor})
into Eq.~(\ref{eq:emissivity}), the emissivity separates into
spatial and energy factors:
\begin{equation}
  \varepsilon(E_{\gamma}, \bm{x}) =\,\,
  \underbrace{n_{\mathrm{gas}}(\bm{x})\,u_{\CR}(\bm{x})}_{\text{spatial}}\qquad 
  \;\times\;
  \underbrace{c\!\int\!
              \frac{d\sigma_{pp}}{dE_{\gamma}}\,
              \frac{\widetilde{q}(E_{p})}{E_{p}}\,dE_{p}}_{\text{energy (per unit CR energy per target)}}.
  \label{eq:eps_factor}
\end{equation}
The spatial factor is now the product of two energy densities
per target
In this fiducial calculation we
take the target gas density as spatially uniform, representing
the hot diffuse component of the inner-Galaxy ISM in the
Fermi-bubble cavity,
\begin{equation}
  n_{\mathrm{gas}}(\bm{x}) \;\approx\; n_{\mathrm{eff}}
  \;\simeq\; 0.03\,\mathrm{cm^{-3}}
  \qquad
  (n_{\mathrm{eff}} \in [0.01, 0.05]\,\mathrm{cm^{-3}}),
  \label{eq:neff}
\end{equation}
where $n_{\rm eff}$ denotes the target hydrogen (nucleon)
density entering the $pp$ interaction
The fiducial value is of order the
Galactic Center normalisation of the inner-Galaxy hot-halo
model~\cite{ZhangGuo2020}. The lower
bound $0.01\,\mathrm{cm^{-3}}$ is appropriate to the depleted
Fermi-bubble cavity, where the emission measure drops by
$\simeq 50\%$~\cite{Kataoka2013}; the upper bound
$0.05\,\mathrm{cm^{-3}}$ brackets the denser inner-bulge end.
The spatial structure of $\varepsilon$ is therefore controlled
by $u_{\CR}(\bm{x})$ alone; the dense molecular and atomic gas
of the bar is restored in Sec.~\ref{sec:triaxial} through
$n_{\mathrm{gas}}(\bm{x}) = n_{\mathrm{bar}}(\bm{x}) +
n_{\mathrm{halo}}$, with $n_{\mathrm{halo}} \equiv
n_{\mathrm{eff}}$, where the triaxial bar gives rise to the
longitudinal asymmetry of the GCE.

\subsection{Isotropic CR cloud approximation}
\label{ssec:factorisation}

The precession-averaged source pattern $P(\vartheta)$ is
concentrated in the polar band
$15^{\circ} \le \vartheta \le 55^{\circ}$ and its southern
conjugate, with $P(\vartheta) = 0$ near both the Galactic
poles and the plane. Taken at face value this would predict
an annular projected morphology, in tension with the
approximately spherical GCE template
fits~\cite{DiMauro2021,Calore2015}.

Isotropic diffusion resolves this through the centrifugal
suppression of higher Legendre multipoles in the radial Green
function. For an injected pattern
$P(\vartheta) = \sum_{\ell} a_{\ell} P_{\ell}(\cos\vartheta)$
and isotropic $D$, each multipole evolves with a radial mode
carrying a $-\ell(\ell+1)/r^{2}$ centrifugal barrier. Inside
the diffusion sphere ($r \lesssim \lambda_{\rm bulge}$),
higher-$\ell$ modes are suppressed relative to the monopole
by $(r/2\sqrt{D\tact})^{\ell}$ at leading order, with a
smooth confluent-hypergeometric crossover
(Appendix~\ref{app:multipole}).

For the fiducial $(i_{\tilt},\theta_{c}) =
(35^{\circ},20^{\circ})$ pattern, the source
quadrupole-to-monopole ratio is $a_{2}/a_{0} \simeq 2.31$; at
$r \sim 1\,\kpc$ ($r/\lambda_{\rm bulge} \simeq 0.58$) the
diffusive damping gives $b_{2}/b_{0} \simeq 5\times 10^{-3}$,
so the net quadrupole modulation of $u_{\CR}$ is at the
few-percent level (Table~\ref{tab:multipole_ratios}). The
hexadecapole $a_{4}/a_{0} \simeq -1.12$ is further suppressed
by $(r/\lambda)^{2}$, giving $u_{4}/u_{0} \lesssim 10^{-4}$.

We accordingly retain only the $\ell = 0$ mode,
\begin{equation}
  u_{\CR}(\bm{x}) \;\approx\; u_{\CR}(r),
  \label{eq:nCR_iso}
\end{equation}
with the residual $\lesssim 3\%$ quadrupole subsumed in the
factorisation uncertainty. The radial profile is governed by
the spherically-symmetric diffusion equation with
point-source injection (Sec.~\ref{ssec:Green}). We adopt
$r_{\rm dep} = 100\,\mathrm{pc}$ as the inner LOS cutoff: the
BZ acceleration zone is distributed over the inner
$\sim 100\,\mathrm{pc}$, so the point-source idealisation
breaks down at $r \lesssim r_{\rm dep}$. The condition
$r_{\rm dep}/\lambda_{\rm bulge} \simeq 0.06 \ll 1$ ensures
the shell source acts as a point source in the multipole
analysis. Two limits in which the monopole approximation may
degrade --- slow precession ($\Prec \lesssim 1$) and strongly
anisotropic diffusion --- are not realized in the fiducial
case and are discussed in Sec.~\ref{sec:conclusions}.

\subsection{Two-zone CR transport}
\label{ssec:Green}

The radial CR energy density $u_{\CR}(r)$ is obtained by
solving the spherically-symmetric two-zone diffusion problem
directly. We summarise the governing equation, the two-zone
diffusion coefficient, and the post-shutoff corrections here;
the numerical scheme is described in Sec.~\ref{ssec:numerics},
and the closed-form quasi-stationary solution used to validate
the numerics is given in Appendix~\ref{app:greenbox}.

\paragraph{Diffusion equation.}
We model the CR source as a delta function at the GC. The
deposition scale $r_{\rm dep} = 100\,\mathrm{pc}$ is much
smaller than the present-day $\lambda_{\rm bulge} \simeq
2\,\kpc$, so the source is effectively pointlike for radial
transport. Acceleration on inner scales proceeds through
reconnection in the BZ flow, internal shocks, and shear
acceleration along the jet column~\cite{Kimura2018}; the
mechanism does not enter the morphology beyond setting the
source size. We do not consider alternative deposition
geometries (jet termination shock, volume-filling cocoon),
appropriate for the bubble morphology~\cite{Yang2022}. CR
transport in the bulge is governed by
\begin{equation}
  \frac{\partial u_{\CR}}{\partial t}
  = \nabla \!\cdot\! \bigl[D(\bm{x})\,\nabla u_{\CR}\bigr]
    + Q(\bm{x}, t),
  \label{eq:diffusion_pde}
\end{equation}
with energy losses neglected since $t_{pp} \gg \tact$
(Sec.~\ref{ssec:budget}). For the radial calculation,
\begin{equation}
  Q(\bm{x}, t) = Q_{0}\,\delta^{3}(\bm{x})\,
                 \Theta(t)\,\Theta(\tact - t),
  \label{eq:Q_source}
\end{equation}
with $Q_{0}$ the constant injection rate, fixed by the energy
budget through $Q_{0}\,\tact = E_{\CR}$ by construction.

\paragraph{Two-zone structure.}
Three nested radial scales control transport, each anchored
physically and not fit: the deposition cutoff
$r_{\rm dep} = 100\,\mathrm{pc}$; the CMZ--bulge boundary
$R_{b} = 200\,\mathrm{pc}$ (the molecular-gas extent over
which HESS measures the $\propto 1/r$ CR
profile~\cite{HESS2016}); and the bulge confinement radius
$R_{\rm bulge} = 2\,\kpc$ (leaky-box escape,
Sec.~\ref{ssec:spectrum}). The dense, magnetized CMZ
($r_{\rm dep} \le r \le R_{b}$) has suppressed diffusion
$D_{\rm CMZ} = 10^{27}\,\mathrm{cm^{2}\,s^{-1}}$, while the
surrounding bulge has $D_{\rm bulge} = 3 \times
10^{28}\,\mathrm{cm^{2}\,s^{-1}}$:
\begin{equation}
  D(r) =
  \begin{cases}
    D_{\mathrm{CMZ}}    & r < R_{b},\\[2pt]
    D_{\mathrm{bulge}}  & r \ge R_{b}.
  \end{cases}
  \label{eq:2zone}
\end{equation}
The suppressed inner value is characteristic of CRs in dense
magnetized gas~\cite{GabiciAharonianBlasi2007} and of dedicated
CMZ transport models~\cite{Scherer2023}; the bulge value
matches plane propagation models~\cite{Gaggero2015}. The
morphology fit changes by $\lesssim 20\%$ for
$D_{\rm CMZ} \in [10^{27}, 10^{28}]\,\mathrm{cm^{2}\,s^{-1}}$.
The associated diffusion lengths
$\lambda_{i} \equiv \sqrt{4 D_{i} \tact}$ are
$\lambda_{\rm CMZ} \simeq 0.32\,\kpc$ and
$\lambda_{\rm bulge} \simeq 1.73\,\kpc$ at
$\tact = 7.5\,\Myr$ (active-phase values; the broadened
present-day $\lambda_{\rm bulge}$ is larger by the factor
below). Because the CMZ crossing time
$R_b^2/6D_{\rm CMZ} \simeq 2\,\Myr$ is short compared to
$\tact$, the inner zone reaches a flow-through steady state
well before shutoff; only $\sim 5\%$ of the injected CR energy
resides inside $R_b$ at the present day
(Sec.~\ref{ssec:numerics}).

\paragraph{Post-shutoff corrections.}
Two physically distinct corrections
convert the active-phase configuration to the present day:
\begin{itemize}
  \item {Continued diffusion.} With injection switched
  off, the existing cloud keeps spreading for $T_{\rm off}$.
  This is captured exactly by evolving
  Eq.~(\ref{eq:diffusion_pde}) to the full
  $T_{\rm eff} = \tact + T_{\rm off} = 10.1\,\Myr$ with $Q = 0$
  for $t > \tact$, broadening
  $\lambda_{\rm bulge}\to2.0\,\kpc$. The evolution is
  number- and energy-conserving by construction of the
  finite-volume scheme.
  \item {Adiabatic expansion.} The bubble expands by
  $x = (T_{\rm eff}/\tact)^{4/7} \simeq 1.19$, shifting the
  CR spectrum down in energy. For $N(E)\propto E^{-\alpha}$
  the differential flux at fixed photon energy is reduced by
  $x^{1-\alpha} \simeq 0.79$ (a $\sim 21\%$ reduction at
  $\alpha = 2.4$); the spectral slope is unchanged and the
  cutoff shifts $100 \to 84\,\TeV$, irrelevant in the GeV
  band.
\end{itemize}
The escape-driven spectral shaping (
Sec.~\ref{ssec:spectrum}) freezes at shutoff, while the
spatial cloud continues to spread; this deliberate asymmetry
reflects that injection --- and hence escape shaping --- stops
at $\tact$. The continued-diffusion correction is built into
the numerical evolution; the scalar adiabatic factor multiplies
the final spectrum and brightness
(Sec.~\ref{ssec:numerics}).

\subsection{Energy-resolved spectrum: leaky-box treatment}
\label{ssec:spectrum}

The in-bulge spectrum is shaped by the competition between
continuous injection over $\tact$ and energy-dependent
diffusive escape. Following the leaky-box
treatment~\cite{Jouvin2020}, the effective in-bulge residence
time is
\begin{equation}
    \tau_{\rm eff}(E_p) \;=\;
    \min\!\left[\tact,\;
      \frac{R_{\rm bulge}^{2}}{6\,D_{0}\,(E_p/E_{0})^{\delta}}
    \right],
  \label{eq:tau_eff}
\end{equation}
with $R_{\rm bulge} = 2\,\kpc$, $D_{0} = 3\times
10^{28}\,\mathrm{cm^{2}\,s^{-1}}$, $E_{0} = 1\,\GeV$, and
$\delta = 0.5$ the quasi-linear diffusion index expected for a
Kraichnan turbulence spectrum~\cite{Kraichnan1965}.
The in-bulge differential number spectrum is the leaky-box
steady state, injection rate times residence time,
\begin{equation}
  N(E_p) \;=\;
    E_p^{-\alpha}\,
    \exp\!\left(-E_p/E_p^{\rm cut}\right)\,
    \tau_{\rm eff}(E_p),
  \label{eq:N_inbulge}
\end{equation}
with $\alpha = 2.4$ a fiducial injected index consistent with
non-thermal acceleration in magnetized jets and
$E_p^{\rm cut} = 100\,\TeV$ a conservative cutoff well below
the Hillas ceiling computed below. The residence time
$\tau_{\rm eff}$ Eq.~(\ref{eq:tau_eff}) produces a break at
\begin{equation}
  E_p^{\star} \;\equiv\;
    E_{0}\!\left(\frac{R_{\rm bulge}^{2}}{6\,D_{0}\,\tact}
    \right)^{1/\delta}
    \;\simeq\; 0.8\,\GeV,
  \label{eq:Epstar}
\end{equation}
above which $\tau_{\rm eff} \propto E_p^{-\delta}$ and the
spectrum steepens from $\alpha$ to $\alpha + \delta = 2.9$.
The energy-normalized shape passed to the radiative
calculation is
\begin{equation}
  \widetilde{q}(E_p) \;=\;
    \frac{N(E_p)}
         {\displaystyle\int_{E_p^{\min}}^{E_p^{\rm cut,int}}\!
          E_p'\,N(E_p')\,dE_p'},
  \qquad
  \int_{E_p^{\min}}^{E_p^{\rm cut,int}}\! E_p\,\widetilde{q}(E_p)\,dE_p = 1,
  \label{eq:q_norm}
\end{equation}
with $E_p^{\min} = 1\,\GeV$ and the integration extended to
$E_p^{\rm cut,int} = 10\,\PeV$ (two orders beyond
$E_p^{\rm cut}$ for convergence of the suppressed tail), so
that the proton population $\widetilde{q}$ represents carries
unit total energy (Sec.~\ref{ssec:numerics}).

\paragraph{Maximum proton energy.}
A magnetized region can accelerate a proton only while its
gyroradius remains smaller than the region itself. We therefore
verify that each acceleration site can confine protons to at
least the fiducial cutoff $E_p^{\rm cut} = 100\,\TeV$ by
comparing the proton Larmor radius with the region size $L$ ---
taken transverse to particle confinement: the launching-zone
extent at site~(i), the jet transverse radius $r_\perp$ at
site~(ii), and the shock scale at site~(iii).
The Larmor radius (orbit size) for a relativistic proton in a
uniform field is
\begin{equation}
  r_{L}(E_{p}) \;=\; \frac{E_{p}}{eB}
  \;\simeq\; 1.08 \times 10^{-6}\,\mathrm{pc}
  \left(\frac{E_{p}}{1\,\mathrm{PeV}}\right)
  \left(\frac{B}{1\,\mathrm{G}}\right)^{-1}.
  \label{eq:rL}
\end{equation}
Confinement requires $r_L \le L$ in a magnetized region of
size $L$, and the maximum confinable energy with characteristic
flow velocity $\beta_{\mathrm{sh}}c$ is the Hillas
energy~\cite{Hillas1984}
\begin{equation}
  E_{p}^{\max} \;=\; Z\,e\,B\,L\,\beta_{\mathrm{sh}}.
  \label{eq:hillas}
\end{equation}
We evaluate Eq.~(\ref{eq:hillas}) at three relevant sites of
the system. (i) Near the black hole horizon, $B \sim 2\times
10^{4}\,\mathrm{G}$ within the active jet-launching zone
$L \sim 10\,r_{g}$, and $\beta_{\rm sh} \sim 0.1$
characteristic of the BZ outflow at the fast-magnetosonic
surface, giving $E_{p}^{\max} \simeq 3.8\,\mathrm{EeV}$.
(ii) Along the jet at $s \simeq 0.1\,\kpc$, the equipartition
field is $B_{jet} \simeq 100\,\mu\mathrm{G}$ Eq.~(\ref{eq:Bjet})
and the transverse scale is $r_{\perp} \simeq 9\,\mathrm{pc}$,
with $\beta_{\rm sh} \simeq 0.1$ characteristic of
relativistic jet shear/reconnection
acceleration~\cite{Rieger2019},
yielding $E_{p}^{\max} \simeq 80\,\mathrm{PeV}$.
(iii) At the post-shutoff Fermi/eROSITA bubble forward shock,
$B_{\wall} \sim 4\,\mu\mathrm{G}$ (total wall field;
Sec.~\ref{ssec:Bfield}), $L \sim 5\,\kpc$, and
$v_{\rm sh} \equiv dR_{\parallel}/dt \simeq 390\,\mathrm{km/s}$
(from the present-day $t^{4/7}$ coasting expansion;
Sec.~\ref{ssec:jet_single_source}) give
$\beta_{\rm sh} \simeq 1.3\times 10^{-3}$
and $E_{p}^{\max} \simeq 22\,\mathrm{PeV}$. All these sites  accommodate our adopted spectral
cutoff $E_{p}^{\rm cut} = 100\,\TeV$ by at least two orders of
magnitude, so cosmic-ray acceleration is not energy-limited
within the fiducial model; the cutoff reflects our conservative
spectral choice rather than a confinement bound. The BZ-jet
model is consistent with proton injection up to
$\gtrsim 100\,\TeV$ at the jet base and along the collimated
jet, and subsequent diffusion populates the GCE region at
$E_p \lesssim E_p^{\rm cut}$.

\paragraph{Optical depth to pair production.}
Finally, the inner Galaxy is optically thin to the GCE photons,
so no attenuation correction is applied. For a
$100\,\mathrm{GeV}$ gamma ray the head-on threshold requires a
target photon $E_{\rm t} \gtrsim m_e^2 c^4/E_\gamma \simeq
2.6\,\mathrm{eV}$. With the number density of interstellar
radiation field (ISRF) photons above the pair-production
threshold $n_{\rm ISRF} \sim 1\,\mathrm{cm}^{-3}$~\cite{Moskalenko2006}
and adopting the peak cross-section $\sigma_{\gamma\gamma}
\simeq 1.2 \times 10^{-25}\,\mathrm{cm}^{2}$~\cite{GouldSchreder1967}
as a conservative bound,
\begin{equation}
  \tau_{\gamma\gamma} = \int_{0}^{d_{\odot}}\!
                       n_{\mathrm{ISRF}}\,
                       \sigma_{\gamma\gamma}\,ds
                       \;\simeq\; 0.003 .
\end{equation}
A factor-of-ten higher $n_{\rm ISRF}$ gives
$\tau_{\gamma\gamma} \sim 0.03$, still $\ll 1$, confirming the
medium is optically thin in the GeV band.
\subsection{Surface brightness}
\label{ssec:Ib}

The observed photon flux per steradian along a line of sight
at $(\ell, b)$ is
\begin{equation}
  I(\theta, E_{\gamma}) =
  \frac{1}{4\pi}\int_{0}^{S_{\max}}\!
    \varepsilon\bigl(E_{\gamma},\,\bm{x}(s; \ell, b)\bigr)\,ds,
  \label{eq:dNdO}
\end{equation}
in $\mathrm{ph\,cm^{-2}\,s^{-1}\,GeV^{-1}\,sr^{-1}}$, with
$S_{\max} = 16\,\kpc$ (matching the numerical
integration endpoint of Sec.~\ref{ssec:numerics}); the
injected energy density falls off on the scale
$\lambda_{\rm bulge} \simeq 2\,\kpc$ and is negligible beyond a
few $\lambda_{\rm bulge}$, so the truncation is effectively
exact for the morphology fit. For a point at distance $s$ along
$\ell = 0$, latitude $b$,
\begin{align}
 \rho_{\mathrm{GC}}(s, b) = |d_{\odot} - s\cos b|,\qquad
  z_{\mathrm{GC}}(s, b) = s\sin b,\\
 r(s, b) = \sqrt{\rho_{\rm GC}^{2} + z_{\rm GC}^{2}},\qquad
  \vartheta(s, b) = \cos^{-1}\!
                     \bigl[z_{\rm GC}(s, b)/r(s, b)\bigr]
                   \in [0, \pi].
  \label{eq:r_gc}
\end{align}
The radial coordinate $r$ is the only Galactocentric input to
the spherically-symmetric $u_{\CR}(r)$; $\vartheta$ is
retained for the multipole analysis
(Appendix~\ref{app:multipole}) and the triaxial calculation
(Sec.~\ref{sec:triaxial}) but does not enter the fiducial LOS
integral. Because the CR and gas distributions are
azimuthally symmetric, the surface brightness depends only on
$\theta$; for $\ell = 0$, $\theta = |b|$.

Substituting Eq.~(\ref{eq:eps_factor}) into
Eq.~(\ref{eq:dNdO}) with the isotropic cloud and uniform gas,
\begin{equation}
  I(\theta, E_{\gamma})
  = \frac{1}{4\pi}\,f_{\rm shape}(E_{\gamma})\,
    \mathcal{L}(\theta),
  \label{eq:Ib}
\end{equation}
where
\begin{equation}
  \mathcal{L}(\theta) \;\equiv\;
    \int_{0}^{S_{\max}}\!
    u_{\CR}\bigl(r(s, \theta)\bigr)\,
    \Theta\!\bigl(r(s,\theta) - r_{\rm dep}\bigr)\,ds,
  \label{eq:LtAxis}
\end{equation}
with the inner $r < r_{\rm dep} =
100\,\mathrm{pc}$ excluded. The observed morphology is
controlled primarily by the projected radial profile
$u_{\CR}(r)$, not by the proton spectrum. The energy prefactor
is
\begin{equation}
  f_{\rm shape}(E_{\gamma})
  = n_{\rm eff}\,c\!
    \int_{E_{\rm th}(E_{\gamma})}^{\infty}\!
    \frac{d\sigma_{pp}}{dE_{\gamma}}(E_{p}, E_{\gamma})\,
    \frac{\widetilde{q}(E_{p})}{E_{p}}\,dE_{p},
  \label{eq:fshape_def}
\end{equation}
with $E_{\rm th}(E_{\gamma})$
the minimum proton energy producing a photon $E_{\gamma}$; the
lower limit is set by the kinematics in the Kafexhiu et
al.~\cite{Kafexhiu2014} parametrisation.
Galactic-disc emission is not included.
For comparison with the
measured GCE radial profile, we integrate
Eq.~(\ref{eq:Ib}) over the $1$--$10\,\GeV$ band,
\begin{equation}
  \frac{dN}{d\Omega} \;\equiv\;
  \int_{1\,\GeV}^{10\,\GeV}\!
    E_{\gamma}\, I(\theta, E_{\gamma})\, dE_{\gamma}
  \label{eq:dNdOmega_band}
\end{equation}
 following the
 notation and convention of Ref.~\cite{DiMauro2021}.
\subsection{The GCE spectrum}
\label{ssec:spectrum_recipe}

The region-averaged spectrum compared with the GCE measurement
follows from the same emissivity by averaging the intensity,
Eq.~(\ref{eq:dNdO}), over the region of interest (ROI),
\begin{equation}
  \Bigl\langle \frac{dN}{dE_\gamma} \Bigr\rangle_{\rm ROI}
  = \frac{\int_{\rm ROI} I(\theta, E_{\gamma})\,d\Omega}
         {\int_{\rm ROI} d\Omega}
  = \frac{f_{\rm shape}(E_{\gamma})}{4\pi}\,
    \langle\mathcal{L}\rangle_{\rm ROI},
  \label{eq:spec_roi}
\end{equation}
with $\langle\mathcal{L}\rangle_{\rm ROI}$ the
solid-angle--weighted mean of Eq.~(\ref{eq:LtAxis}) over
$\theta < 10^{\circ}$, matching the region of the spectral
measurement of Ref.~\cite{DiMauro2021}. The plotted quantity
is $E_\gamma^2\,\langle dN/dE_\gamma\rangle_{\rm ROI}$,
multiplied by the post-shutoff adiabatic factor
$x^{1-\alpha}\simeq0.79$. The resulting spectrum and its
comparison with the GCE measurement are presented in
Sec.~\ref{sec:chi2}.

\subsection{Numerical implementation}
\label{ssec:numerics}

\paragraph{Transport.}
The two-zone diffusion problem, Eq.~(\ref{eq:diffusion_pde}),
with step-function diffusivity Eq.~(\ref{eq:2zone}), is solved
using a conservative finite-volume scheme on a radial grid with
$2\,\mathrm{pc}$ spacing extending to $8\,\kpc$. The diffusion
coefficient is defined at cell interfaces using a harmonic mean
across the $R_b$ discontinuity, which ensures stable numerical
fluxes and enforces continuity of $D\,\partial_r u_{\CR}$ at the
CMZ--bulge interface at the discrete level. The source is injected uniformly within the innermost
$30\,\mathrm{pc}$ (well inside the characteristic deposition
scale $r_{\rm dep}$) at constant rate $Q_0$ during
$0 < t < \tact$. After $t=\tact$, the system is evolved with
$Q=0$ up to $T_{\rm eff} = 10.1\,\Myr$, corresponding to
post-injection diffusion. This procedure is equivalent to
linear superposition of Green-function solutions of the diffusion
equation, which we have verified numerically by comparison with
the full time-dependent evolution. The scheme conserves total CR energy to better than a few parts
in $10^{6}$. Resolution tests comparing $4\,\mathrm{pc}$ and
$2\,\mathrm{pc}$ grids show convergence of the radial profile at
the $\lesssim 0.3\%$ level (L2 norm) for $r \gtrsim 50\,\mathrm{pc}$.
The resulting CR profile is normalized using
Eq.~(\ref{eq:ECR_def}). At the present epoch, approximately
$\sim 5\%$ of the total CR energy lies within $R_b$, and
$\sim 33\%$ within $1\,\kpc$. The CR energy density at
$500\,\mathrm{pc}$ is $u_{\CR} \simeq 1.3\,\mathrm{eV\,cm^{-3}}$,
consistent with typical Galactic values. As an additional cross-check, the numerical solution is compared
with the quasi-stationary analytic approximation of
Appendix~\ref{app:greenbox}, which reproduces the numerical
profile to within $\lesssim 25\%$ over the range
$0.4 \lesssim r \lesssim 3\,\kpc$. This agreement provides an
independent validation of the diffusion implementation.

\paragraph{Spectrum and radiation.}
The in-bulge proton spectrum
Eq.~\eqref{eq:N_inbulge} is evaluated on
a logarithmic grid of 500 energy points spanning
$1\,\GeV$ to $10\,\PeV$, ensuring numerical convergence of the
high-energy tail. The spectrum is normalized using
Eq.~(\ref{eq:q_norm}). The hadronic gamma-ray yield $f_{\rm shape}(E_\gamma)$ is computed
using \texttt{naima}~\cite{Zabalza2015naima} with the
\texttt{PionDecay} module and the parameterisation of
Kafexhiu et al.~\cite{Kafexhiu2014}, assuming a target density
$n = 1\,\mathrm{cm^{-3}}$. The physical gas density and CR energy
budget enter separately through the spatial weighting factor
$\mathcal{L}(\theta)$, as described in Eq.~(\ref{eq:eps_factor}).
No radiative cooling of protons is included, as hadronic loss
timescales are much longer than diffusion timescales in the
parameter regime considered.

\paragraph{Projection.}
Line-of-sight integrals in Eqs.~(\ref{eq:Ib}) and
(\ref{eq:LtAxis}) are evaluated using a segmented quadrature
scheme consisting of 30 logarithmic points in the inner region,
250 uniform points across the Galactic Center region, and 50
logarithmic points at large radii up to $16\,\kpc$. Tests with
increased sampling show stability of the resulting intensity
profiles. Azimuthal averaging is performed using 12 uniformly
spaced samples in $\phi$. The present-day spectrum includes a global adiabatic correction
factor $x^{1-\alpha} \simeq 0.79$, applied uniformly to the
emitted flux. This factor represents the effective energy scaling
of the propagated CR population in the adopted one-zone
approximation. The gas density used in the spectral calculation is
$n_{\rm eff} = 0.03\,\mathrm{cm^{-3}}$, while the asymmetry
analysis uses the bar-plus-halo gas model of
Eq.~(\ref{eq:gas_bar_plus_halo}).

\paragraph{Fiducial parameters.}
All fiducial parameters and derived quantities are summarized in
Table~\ref{tab:params} for reference.
\begin{table}[!htbp]
\centering
\caption{Fiducial model parameters and derived quantities.
Entries marked \checkmark{} are
direct inputs to the numerical implementation.}
\label{tab:params}
\small
\setlength{\tabcolsep}{4pt}
\begin{tabular}{@{}llll@{}}
\toprule
Quantity & Symbol & Fiducial value & Input \\
\midrule
\multicolumn{4}{l}{\textit{Black hole and accretion}}\\
BH mass                     & $\MBH$              & $4.15\times10^{6}\,\Msun$ &  \\
BH spin                     & $\astar$            & 0.9 &  \\
Eddington fraction          & $\mdot$             & $1\times10^{-5}$ &  \\
BZ efficiency               & $\eta_{\BZ}$        & $\simeq 0.9$ &  \\
BZ jet power                & $P_{\BZ}$           & $5.23\times10^{40}\,\ergs$ & \checkmark \\
\addlinespace[2pt]
\multicolumn{4}{l}{\textit{Magnetic field}}\\
Horizon field (active)      & $B_{H}^{(0)}$       & $2\times10^{4}\,\mathrm{G}$ &  \\
\addlinespace[2pt]
\multicolumn{4}{l}{\textit{Geometry and timing}}\\
Spin-axis tilt              & $i_{\tilt}$         & $35^{\circ}$ &  \\
BZ cone half-angle          & $\theta_{c}$        & $20^{\circ}$ &  \\
Active duration             & $\tact$             & $7.5\,\Myr$ & \checkmark \\
Time since shutoff          & $T_{\mathrm{off}}$  & $\sim 2.6\,\Myr$ &  \\
Precession period           & $T_{\mathrm{prec}}$ & $\lesssim 1.5\,\Myr$ &  \\
Number of LT cycles         & $\Prec$             & $\gtrsim 5$ &  \\
Warp radius                 & $r_{\mathrm{warp}}$ & $\simeq 9000\,r_{g}$ &  \\
\addlinespace[2pt]
\multicolumn{4}{l}{\textit{Cosmic-ray transport and gamma-rays}}\\
CR injection efficiency     & $\xi_{\CR}$         & $\simeq 0.05$ & \checkmark \\
DSA spectral index          & $\alpha$            & $2.4$ & \checkmark \\
Diffusion energy index      & $\delta$            & $0.5$ & \checkmark \\
Proton cutoff energy        & $E_{p}^{\rm cut}$   & $100\,\TeV$ & \checkmark \\
Effective gas density       & $n_{\rm eff}$       & $0.03\,\mathrm{cm^{-3}}$ & \checkmark \\
$D$ (CMZ)                   & $D_{\rm CMZ}$       & $10^{27}\,\mathrm{cm^{2}\,s^{-1}}$ & \checkmark \\
$D$ (bulge)                 & $D_{\rm bulge}$     & $3\times 10^{28}\,\mathrm{cm^{2}\,s^{-1}}$ & \checkmark \\
Diffusion length (bulge)    & $\lambda_{\rm bulge}$ & $1.73\,\kpc$ ($\sim 12^{\circ}$) &  \\
Diffusion length (CMZ)      & $\lambda_{\rm CMZ}$ & $0.32\,\kpc$ ($\sim 2.2^{\circ}$) &  \\
Deposition cutoff           & $r_{\rm dep}$       & $100\,\mathrm{pc}$ & \checkmark \\
CMZ outer boundary          & $R_b$               & $200\,\mathrm{pc}$ & \checkmark \\
Bulge confinement radius    & $R_{\rm bulge}$     & $2\,\kpc$ & \checkmark \\
Total jet energy            & $E_{\mathrm{jet}}$  & $\simeq 1.2\times10^{55}\,\erg$ &  \\
Total CR energy             & $E_{\CR}$           & $\simeq 6.2\times10^{53}\,\erg$ &  \\
\bottomrule
\end{tabular}
\end{table}
\paragraph{Uncertainties and limitations}

The factorization $I = \mathcal{L}(\theta)\,
f_{\rm shape}(E_{\gamma})/(4\pi)$ evaluates
$\mathcal{L}(\theta)$ at the GeV-scale diffusion coefficient
while including the energy dependence of escape only in
$f_{\rm shape}$ via $\tau_{\rm eff}(E_p)$. A fully
self-consistent calculation would evaluate
$\mathcal{L}(\theta, E_p)$ at each energy, since
$\lambda_{\rm bulge}(E_p)$ grows from $1.7\,\kpc$ at $1\,\GeV$
to $5.5\,\kpc$ at $100\,\GeV$ --- a factor $\sim 3$. This
factorized treatment is standard~\cite{Jouvin2020,Cesarsky1980,
Strong2007}; the resulting morphology uncertainty is expected
at the $\mathcal{O}(10$--$30\%)$ level. This is
separate from, and comparable to, the two post-shutoff
corrections of Sec.~\ref{ssec:Green} (adiabatic flux
$\sim 21\%$, diffusion broadening $\sim 16\%$). A fully
energy-resolved two-zone treatment is deferred to future work.

\section{Galactic Spectrum and Brightness}
\label{sec:chi2}

\begin{figure}[!htbp]
\centering
\includegraphics[width=12cm]{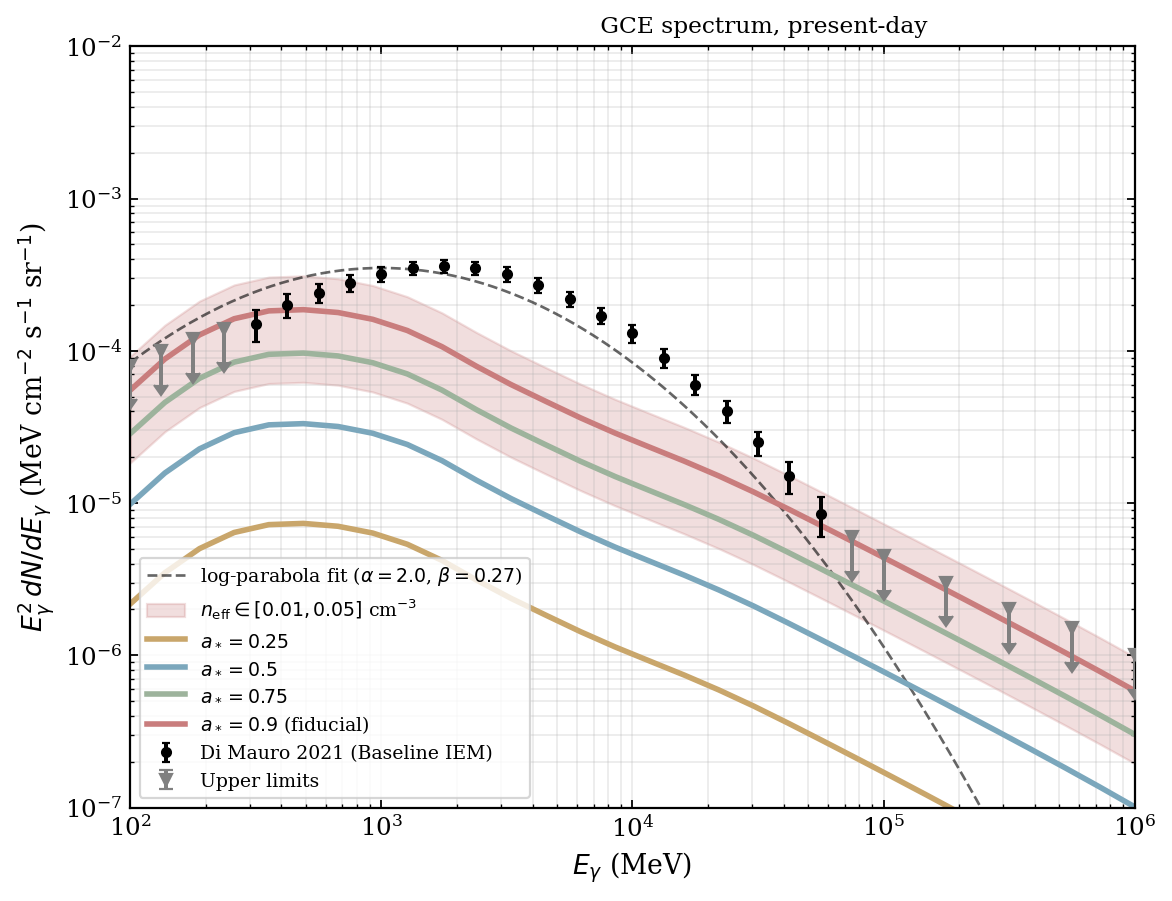}
\caption{Hadronic GCE spectrum integrated over
$\theta < 10^{\circ}$ from the precessing BZ jet of \SgrA{},
for four spins $\astar \in \{0.25, 0.5, 0.75, 0.9\}$ (red band
at fiducial $\astar = 0.9$: $n_{\rm eff} \in [0.01,
0.05]\,\mathrm{cm^{-3}}$), versus the Di Mauro 2021
Baseline-IEM detections~\cite{DiMauro2021} (black points,
$1\sigma$; downward arrows: $95\%$ upper limits). The dashed
grey curve is the published log-parabola fit
($\alpha = 2.0$, $\beta = 0.27$). The model peaks near
$E_{\gamma} \simeq 0.5\,\GeV$. At $\astar = 0.9$ and central
$n_{\rm eff}$ the ROI-averaged amplitude reaches $\sim 105\%$
of the data at $0.3\,\GeV$ and $\sim 50\%$ at $1\,\GeV$,
falling to $\sim 21\%$ at $5$--$10\,\GeV$; at the upper
$n_{\rm eff}$ edge it overshoots the sub-GeV points, which is
what bounds $n_{\rm eff}$ from above. At lower spin the
contribution drops sharply, reaching the percent level at
$\astar \lesssim 0.25$. A hadronic tail at $E_{\gamma} \gtrsim
100\,\GeV$ remains testable with CTA and HAWC.}
\label{fig:dimauro_spectrum}
\end{figure}

\begin{figure}[!htbp]
\centering
\includegraphics[width=12cm]{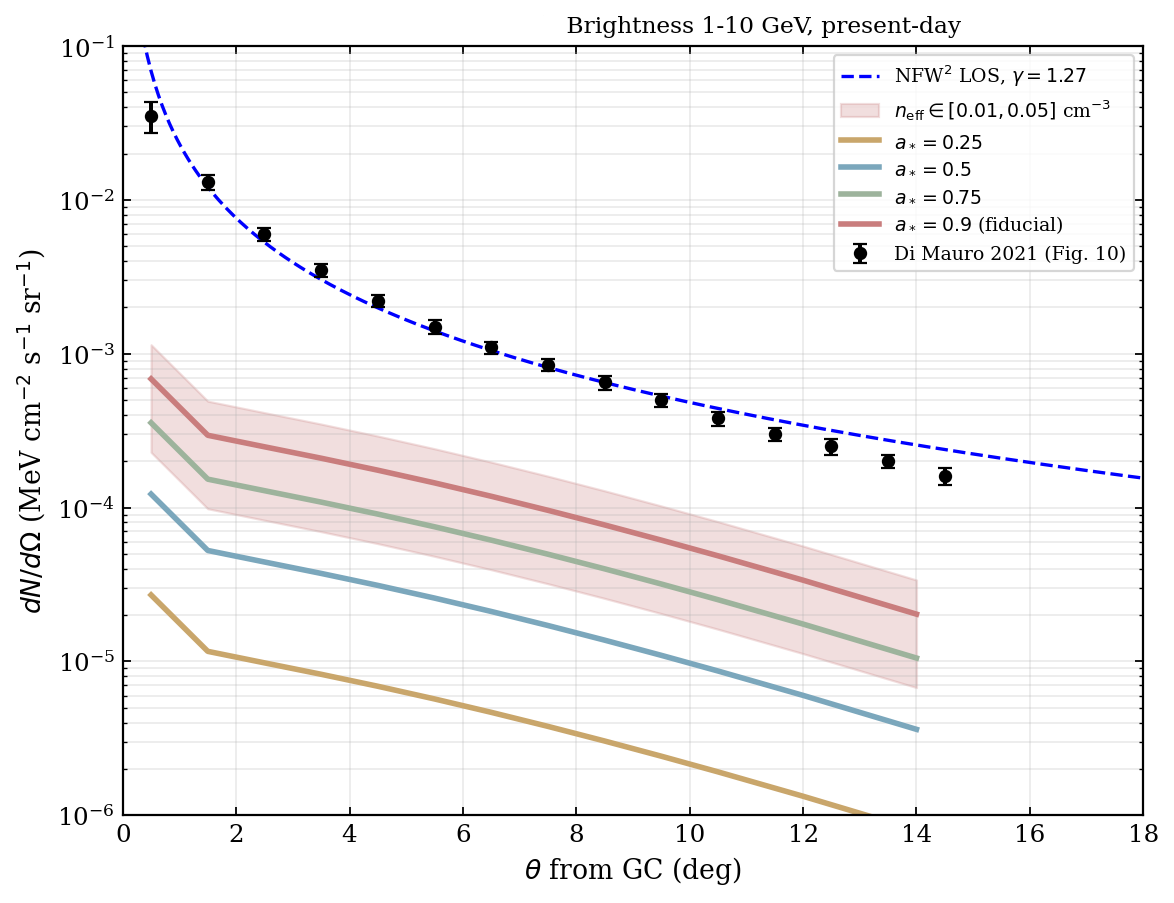}
\caption{Resulting hadronic surface brightness in the
$1$--$10\,\GeV$ band from the precessing BZ jet (four spins;
red band: $n_{\rm eff} \in [0.01, 0.05]\,\mathrm{cm^{-3}}$ at
$\astar = 0.9$), versus the Di Mauro 2021 data (black) and the
best-fit NFW$^{2}$ LOS integral with $\gamma = 1.27$ (dashed
blue). At $\astar = 0.9$ the present-day hadronic floor
(after the post-shutoff adiabatic and diffusion-broadening
corrections of Sec.~\ref{ssec:Green}) is $\sim 2\%$ of the
observed surface brightness at $\theta = 0.5^{\circ}$, rising
to $\sim 12\%$ near $\theta = 9.5^{\circ}$ and
$\sim 13\%$ at $\theta = 14.5^{\circ}$. The shortfall at
small $\theta$ is morphological, not energetic: the
precessing-jet geometry and bulge diffusion smear the CR
injection over $\sim 12^{\circ}$, producing a profile too flat
to match the cuspy NFW$^{2}$ shape.}
\label{fig:dimauro_profile}
\end{figure}

Using the framework of Sec.~\ref{sec:derivation}, we compute
the hadronic GCE surface brightness profile and spectrum from
the precessing BZ jet at fiducial parameters.
Figures~\ref{fig:dimauro_spectrum} and~\ref{fig:dimauro_profile}
compare the calculations against the Di Mauro
2021~\cite{DiMauro2021} measurements for the spectrum
integrated over $\theta < 10^{\circ}$ and the surface
brightness in the $1$--$10\,\GeV$ band.

The computed hadronic spectrum is strongly spin-dependent
through the BZ efficiency $\eta_{\BZ}(\astar)$. At spin
$\astar = 0.9$ and fiducial
$n_{\rm eff} = 0.03\,\mathrm{cm}^{-3}$, the ROI-averaged
amplitude peaks near the $\pi^{0}$-decay bump at $E_{\gamma}
\simeq 0.5\,\GeV$, reaching $\sim 105\%$ of the
detections at $0.3\,\GeV$; at the upper end of the
$n_{\rm eff}$ band it overshoots the sub-GeV points and
crosses the $95\%$ upper limits. Near the GCE spectral peak
($E_{\gamma} \simeq 1.8\,\GeV$) the contribution is
$\sim 33\%$ of the data, falling to $\sim 21\%$
at $5$--$10\,\GeV$ and rising again above $\sim 30\,\GeV$
where the GCE log-parabola steepens. This places the
fiducial high-spin model in mild tension with the sub-GeV
upper limits at the upper end of the $n_{\rm eff}$ band,
favouring either $n_{\rm eff}$ toward the lower, depleted
bubble-cavity value~\cite{Kataoka2013} or a CR
efficiency $\xi_{\CR}$ slightly below the DSA low-end value.
The hadronic tail above $\sim 100\,\GeV$ crosses the GCE
upper limits.

The angular profile in the $1$--$10\,\GeV$ band at $\astar =
0.9$ is substantially flatter than the cuspy NFW$^{2}$
best-fit: at $\theta = 0.5^{\circ}$ the model
contributes $\sim 2\%$ of the GCE, rising
to $\sim 12\%$ at $9.5^{\circ}$ and $\sim 13\%$ at the outer
edge of the ROI ($14.5^{\circ}$). Two features drive the
shallowness: the bulge diffusion length
$\lambda_{\mathrm{bulge}} = 1.73\,\kpc$ smears the central
injection over $\sim 12^{\circ}$, and the precession-averaged
isotropic injection deposits CRs preferentially in the
diffusion-broadened sphere rather than at the cusp. 

The implication is that at high BH spin the BZ-jet
hadronic floor sits at the $\sim 2$--$13\%$ level of the
observed GCE surface brightness and is centrally suppressed;
the remaining $\gtrsim 87\%$ requires a source that is both
more centrally concentrated than the diffusion-broadened
hadronic injection and spectrally softer above $\sim
5\,\GeV$. Both criteria are satisfied by the canonical GCE
interpretations: dark-matter annihilation $\chi\chi \to
b\bar{b}$ with $m_{\chi} \approx 40$--$60\,\GeV$ and a
generalized NFW cusp $\gamma \simeq 1.27$~\cite{DiMauro2021,
Daylan2016,Goodenough2009,Hooper2011}, or an unresolved
millisecond-pulsar population~\cite{Lee2016,Bartels2016,
HolstHooper2023,Kalambay2026} peaking in the bulge. The BZ
floor is therefore an {additional} component that any
complete GCE model must subtract before fitting the dominant
DM or MSP signal; failing to do so biases the inferred DM
cross-section or MSP luminosity at the $\sim 2$--$13\%$
level, comparable to current statistical uncertainties.

The floor amplitude is bounded from below by independent
observations of \SgrA{}'s past activity. Reducing the spin
below $\astar \simeq 0.25$ collapses $\eta_{\BZ} \propto
\Omega_{H}^{2}$ by over an order of magnitude, suppressing the
floor below $\sim 1\%$. Reducing $\mdot$ below
$\sim 10^{-6}\,\dot M_{\Edd}$ would fail to inflate the
bubbles. Reducing $\xi_{\CR}$ below the DSA bound contradicts
PIC/hybrid simulations~\cite{Caprioli2014}. Within these
constraints the morphological floor at $\astar = 0.9$ is
$\sim 2\%$ at minimum and rises to $\sim 30\%$ at the upper
$\xi_{\CR} = 0.3$.

Conversely, three obstacles preclude a BZ-only explanation of
the entire GCE. First, gas density: reproducing the full GCE
peak ($\sim 4 \times 10^{-4}\,\mathrm{MeV\,cm^{-2}\,s^{-1}\,
sr^{-1}}$) would need $n_{\rm eff} \sim 0.3\,\mathrm{cm}^{-3}$
(an order of magnitude above the X-ray value) or $\xi_{\CR}
\sim 1$ (above the DSA maximum) --- neither tenable; indeed
the sub-GeV overshoot already bounds the upper $n_{\rm eff}$.
Second, angular shape: the hadronic profile is too flat for
the NFW$^{2}$ $\gamma = 1.27$ data, because precession plus
diffusion smear the injection --- the cleanest single obstacle.
Third, spectral hardness: the $\pi^{0}$ peak at $\sim
0.5\,\GeV$ lies below the GCE peak at $\sim 1.8\,\GeV$ and is
harder than the log-parabola above $\sim 5\,\GeV$; softening
to $\alpha = 2.6$ would suppress this but reduce the GeV-band
floor.

\section{Triaxial bar gas and longitudinal asymmetry}
\label{sec:triaxial}
The fiducial calculation in Sec.~\ref{sec:derivation} adopts a
spherically-symmetric effective gas density $n_{\rm gas}(\bm{x})
\approx n_{\rm eff} \simeq 0.03\,\mathrm{cm^{-3}}$ representative
of the X-ray-measured hot gas of the inner
bulge~\cite{ZhangGuo2020}. In reality, the
inner-Galaxy interstellar medium contains two distinct
components on the relevant scales: \emph{(i)}~a triaxial
warm-phase contribution that follows the central stellar bar,
with peak density $n_{0}^{\rm bar} \simeq 0.15\,\mathrm{cm^{-3}}$
elongated along the bar major axis at angle $\alpha \simeq
20^{\circ}$ to the Sun--GC line (the low end of the measured
$20$--$30^{\circ}$ range), with the near end at positive
Galactic longitudes~\cite{Wegg2013, Stanek1994}; and
\emph{(ii)}~the hot ionized uniform hot-gas halo. We
replace the uniform ansatz by the additive bar-plus-halo
profile,
\begin{equation}
  n_{\rm gas}(\bm{x}) \;=\; n_{0}^{\rm bar}\,
    \exp\!\Bigl[-\sqrt{(x_{\rm bar}/x_{0})^{2}
                       + (y_{\rm bar}/y_{0})^{2}
                       + (z/z_{0})^{2}}\,\Bigr]
    \;+\; n_{0}^{\rm halo},
  \label{eq:gas_bar_plus_halo}
\end{equation}
with $(x_{\rm bar}, y_{\rm bar}, z_{\rm bar})$ Cartesian
coordinates in the bar-aligned frame:
$x_{\rm bar} = x\cos\alpha + y\sin\alpha$,
$y_{\rm bar} = -x\sin\alpha + y\cos\alpha$,
$z_{\rm bar} = z$, with the bar near-end at positive Galactic
longitude. We adopt fiducial scale lengths $x_{0} = 1.5\,\kpc$,
$y_{0} = z_{0} = 0.4\,\kpc$, representative of the Milky Way
long bar from near-infrared and red-clump-star
surveys~\cite{Wegg2013}. The volume-averaged density within the
inner $1\,\kpc$ cube is $\langle n_{\rm gas} \rangle \simeq
0.057\,\mathrm{cm^{-3}}$, of which
$\sim 53\%$ is the symmetric gas-halo
contribution.

\subsection{The asymmetry decomposition}
\label{ssec:asymmetry_decomp}

The triaxial bar of Eq.~(\ref{eq:gas_bar_plus_halo}), convolved
with the spherically-symmetric two-zone cosmic-ray proton
energy density $u_{\rm CR}(r)$ of
Sec.~\ref{sec:derivation}, breaks the
rotational symmetry about the Galactic Center and
produces a surface brightness that depends on Galactic longitude
$\ell$ (measured in the Galactic plane from the direction of the
Galactic Center) through
Eq.~(\ref{eq:Ib}). Because the LOS integrand separates linearly
in $n_{\rm gas}$, the surface brightness at $b = 0$ decomposes
into bar and gas-halo contributions,
\begin{equation}
  I(\ell, 0) \;=\; \int u_{\rm CR}(r)\,n_{\rm bar}(\bm{x})\,ds
             \;+\; n_{0}^{\rm halo}\!\int u_{\rm CR}(r)\,ds
             \;\equiv\; I_{\rm bar}(\ell)
                       + I_{\rm halo}(|\ell|),
  \label{eq:LOS_decomp}
\end{equation}
in which the gas-halo term is invariant under $\ell \to -\ell$
because $u_{\rm CR}$ is spherically symmetric and $n_{0}^{\rm
halo}$ is uniform. The asymmetry ratio is therefore
\begin{equation}
  R(\ell) \;\equiv\; \frac{I(+\ell, 0)}{I(-\ell, 0)}
          \;=\; \frac{r_{\rm bar}(\ell) + f(\ell)}
                     {1 + f(\ell)},
  \label{eq:R_dilution}
\end{equation}
where $r_{\rm bar}(\ell) \equiv I_{\rm bar}(+\ell) / I_{\rm
bar}(-\ell)$ is the pure-bar asymmetry and $f(\ell) \equiv
I_{\rm halo}(|\ell|) / I_{\rm bar}(-\ell)$ measures the relative
gas-halo contribution on the dimmer side. The asymmetry is
\emph{generated} by the bar gas: $f \to 0$ would give $R = r_{\rm
bar}$. The gas halo \emph{regulates} the amplitude: as $|\ell|$
grows and the LOS at $-\ell$ exits the bar, $r_{\rm bar}(\ell)$
diverges, but $f(\ell)$ grows in step and dilutes $R$ by
$(1 + f)^{-1}$, yielding a finite \emph{peaked} asymmetry where
bar contrast is strongest relative to gas-halo dilution.

A central feature of Eq.~(\ref{eq:R_dilution}) is that $R(\ell)$
is independent of the overall normalisation of $u_{\rm CR}$: the
cosmic-ray budget $E_{\CR}$, the black-hole spin through
$\eta_{\BZ}(\astar)$, the injection efficiency $\xi_{\CR}$, and
the bar peak density $n_{0}^{\rm bar}$ all enter $I(+\ell)$ and
$I(-\ell)$ as common multiplicative factors and cancel in the
ratio. The same cancellation removes the entire energy-dependent
pion-decay yield, since the spectral shape is position-independent
(Sec.~\ref{ssec:emissivity}); the band-integrated $R(\ell)$
therefore depends only on the geometry of $u_{\rm CR}(r)$ and on
the single dimensionless dilution parameter set by $n_{0}^{\rm
halo}$. The asymmetry is thus a purely geometric diagnostic,
insensitive to the most uncertain energetic inputs of the model.

Geometrically, a LOS at Galactic longitude $\ell$ ($b = 0$)
crosses the bar major axis at Galactocentric radius
\begin{equation}
  r_{\rm int}(\pm\ell) \;=\; \frac{d_{\odot}\,\sin|\ell|}
                                  {\sin(\alpha \pm |\ell|)},
  \label{eq:rint}
\end{equation}
where the upper (lower) sign applies for positive (negative)
longitudes. At positive $\ell$ the LOS samples the bar near-end
at smaller $r$, where the inner-zone CR density $u_{\rm CR}(r)$
rises steeply; at negative $\ell$ the same gas density
is encountered at larger $r$, where the CR density is suppressed
by diffusive escape.

\subsection{Asymmetry profile}
\label{ssec:asymmetry_profile}

\begin{figure}[!htbp]
\centering
\includegraphics[width=0.85\linewidth]{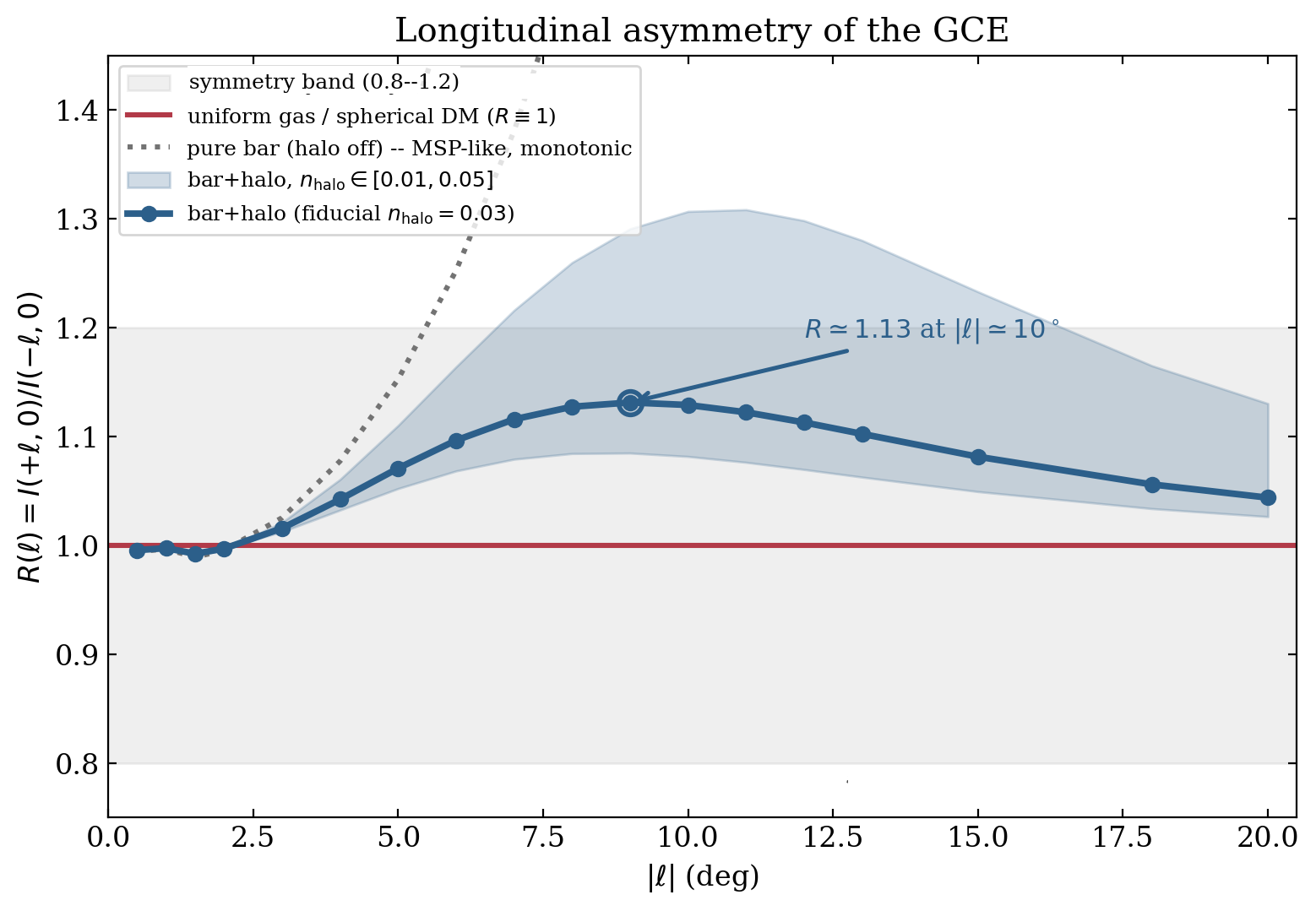}
\caption{Longitudinal asymmetry of the GCE at zero Galactic
latitude, $R(\ell) = I(+\ell, 0)/I(-\ell, 0)$, as a function of
absolute Galactic longitude $|\ell|$. The blue curve is the
fiducial bar-plus-halo result
[$n_{0}^{\rm halo} = 0.03\,\mathrm{cm^{-3}}$,
Eq.~(\ref{eq:gas_bar_plus_halo})], peaking at $R \simeq 1.13$
near $|\ell| \approx 10^{\circ}$ and declining at larger $|\ell|$
as the symmetric gas halo dominates the line-of-sight integral.
The blue band spans the X-ray-allowed halo range
$n_{0}^{\rm halo} \in [0.01, 0.05]\,\mathrm{cm^{-3}}$, within
which the peak amplitude ranges from $R \simeq 1.09$ (denser,
more diluted) to $R \simeq 1.29$ (more rarefied, less diluted);
$R(\ell)$ is independent of black-hole spin and cosmic-ray
budget, which cancel in the ratio. The grey dotted curve is the
pure-bar limit ($n_{0}^{\rm halo} \to 0$), which grows
monotonically and is characteristic of an undiluted
bulge-tracing source. The red line is the spherical dark-matter
null $R \equiv 1$, common to all rotationally-invariant profiles.
The grey horizontal band shows, for orientation, the
sphericity range allowed by Di~Mauro 2021, whose ellipsoidal fit
to the GCE gives a major-to-minor axis ratio of $0.8$--$1.2$
along the Galactic plane~\cite{DiMauro2021}; this is an
ellipticity (plane-vs-vertical elongation) rather than a direct
bound on the longitudinal ratio $R(\ell)$, and we display it only
as an indicative tolerance on departures from $R = 1$. The
fiducial prediction lies well within this range.}
\label{fig:asymmetry}
\end{figure}

Numerical integration of Eq.~(\ref{eq:Ib}) along the LOS at $b =
0$ in the $1$--$10\,\GeV$ band, with the bar-plus-halo gas of
Eq.~(\ref{eq:gas_bar_plus_halo}) replacing the uniform $n_{\rm
gas}$, yields the results shown in Fig.~\ref{fig:asymmetry}. The
asymmetry ratio remains within $\sim 5\%$ of unity for $|\ell|
\lesssim 4^{\circ}$, consistent with the approximate quadrant
symmetry ~\cite{DiMauro2021}. At higher
longitudes, the ratio rises to $R \simeq 1.07$ at $|\ell| =
5^{\circ}$ and reaches its peak value $R \simeq 1.13$ near
$|\ell| = 10^{\circ}$, before gradually declining to $R \simeq
1.08$ at $|\ell| = 15^{\circ}$ as the symmetric gas-halo
contribution becomes dominant. Because the spin and cosmic-ray
budget cancel in the ratio, the sole astrophysical uncertainty on
the amplitude is the halo density: over the X-ray-allowed range
$n_{0}^{\rm halo} \in [0.01, 0.05]\,\mathrm{cm^{-3}}$ the peak
spans $R \simeq 1.09$--$1.29$ (blue band), while the peak location
$|\ell| \approx 10^{\circ}$, set by the bar geometry, is
essentially fixed. The fiducial prediction lies well within the
indicative sphericity range, while the
rarefied-halo edge
($n_{0}^{\rm halo} = 0.01\,\mathrm{cm^{-3}}$, $R \simeq 1.29$)
approaches the upper end of that range. A direct comparison would
require the  analysis to be recast as a longitudinal
$R(\ell)$ measurement rather than an ellipticity, but the trend
suggests that a future longitude-binned analysis could use the
observed symmetry to constrain the inner halo density in this
scenario.

Existing morphological fits already favor bulge-shaped templates
over spherical dark-matter profiles~\cite{Macias2019, Calore2021},
which supports the qualitative picture. A spherically-symmetric annihilating
dark-matter component predicts $I(+\ell, b)/I(-\ell, b) \equiv 1$
for any dark-matter halo profile, since the symmetry under $\ell \to -\ell$
at $b = 0$ is required by rotational invariance of any halo about
its center.

\subsection{Discriminating among GCE scenarios}
\label{ssec:asymmetry_test}

\begin{figure*}[!htbp]
\centering
\includegraphics[width=0.49\textwidth]{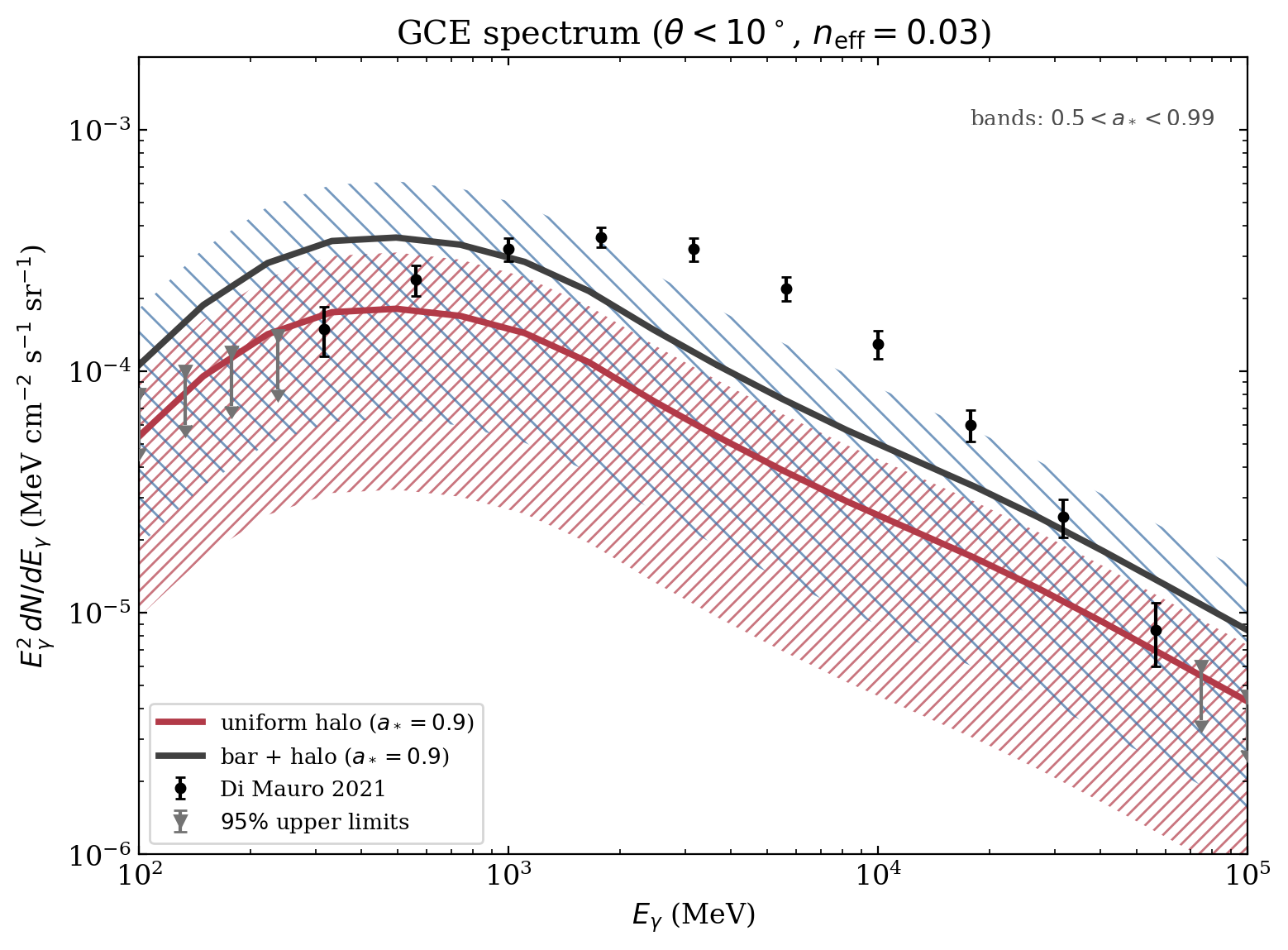}
\hfill
\includegraphics[width=0.49\textwidth]{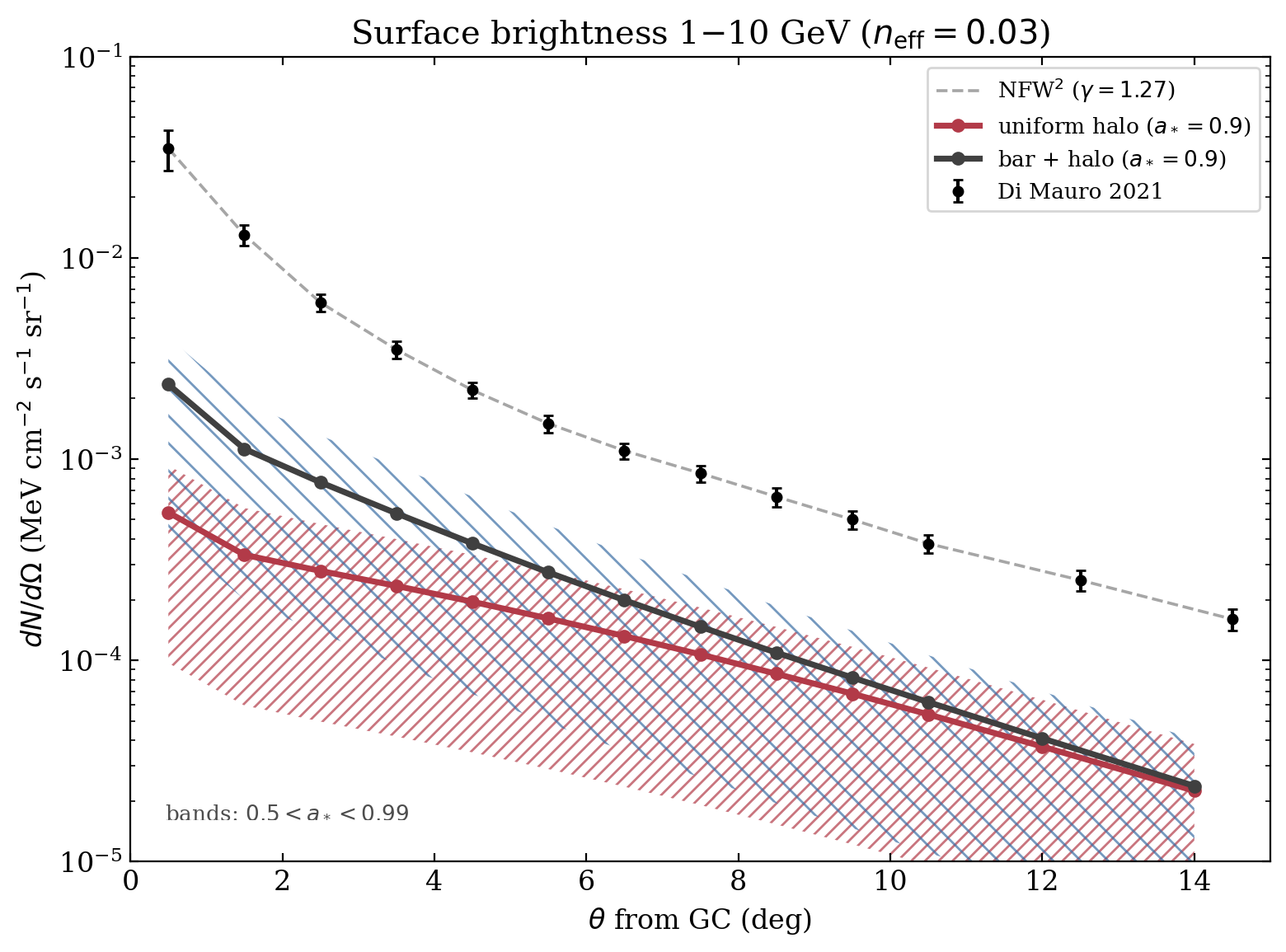}
\caption{Spin dependence of the hadronic GCE floor at fixed halo
density $n_{\rm eff}=0.03\,\mathrm{cm^{-3}}$, comparing the
uniform-halo gas (red) with the triaxial bar-plus-halo gas (dark
solid). \emph{Left:} ROI-averaged spectrum ($\theta<10^\circ$).
\emph{Right:} $1$--$10\,\GeV$ surface-brightness profile. Solid
curves are the fiducial $\astar=0.9$; the hatched bands span
$0.5<\astar<0.99$ (red $/\!/$ uniform, blue $\backslash\backslash$
bar-plus-halo). Black points are the
measurements~\cite{DiMauro2021}; grey downward arrows in the left
panel are their upper limits. The bar roughly doubles the
spectrum amplitude and lifts the inner brightness, but the profile
remains far below the cuspy NFW$^{2}$ template (dashed).}
\label{fig:compare_spin}
\end{figure*}

The bar-plus-halo gas raises the spectrum and steepens the brightness
profile. Neither
change alters the qualitative result --- the hadronic floor stays
sub-dominant and too flat for the cuspy excess --- so the bar's
robust role is the longitudinal asymmetry, where the amplitude
cancels in the ratio, rather than the absolute spectrum or
brightness. The spin dependence is shown in
Fig.~\ref{fig:compare_spin}: varying the spin over
$0.5<\astar<0.99$ shifts the prediction by factors
$0.18$--$1.7$ relative to the fiducial $\astar=0.9$. Comparison
with the \cite{DiMauro2021} upper limits shows that, for the
bar-plus-halo gas at $n_{\rm eff}=0.03\,\mathrm{cm^{-3}}$, the
sub-GeV limits require $\astar\lesssim0.73$;
so a bar gas either drives the preferred spin below
$0.9$ or demands a correspondingly lower $\xi_{\CR}$ or halo
density.

The shape of $R(\ell)$ at zero Galactic latitude offers a useful,
if imperfect, way to distinguish among the leading GCE
candidates, illustrated by the three curves of
Fig.~\ref{fig:asymmetry}. A spherically-symmetric dark-matter
halo predicts $R \equiv 1$
at $b = 0$ for any radial profile (red line), so any detection of asymmetry
above the statistical noise floor is in tension with the
canonical DM interpretation; we note that triaxial DM
halos~\cite{Bryan2013, Prada2019} could introduce a percent-level
asymmetry that partially blurs this distinction. A bulge-tracing
MSP population inherits the bar-elongated morphology of its parent
stellar distribution and produces some asymmetry, but without the
gas-halo dilution of Sec.~\ref{ssec:asymmetry_decomp} it tends to
grow monotonically with $|\ell|$, like the pure-bar limit (grey
dotted), rather than peaking. The hadronic
mechanism developed here suggests a peaked asymmetry maximising
near the bar geometric scale at $|\ell| \approx 10^{\circ}$ before
declining as the symmetric gas-halo contribution dominates; both the
peak location and amplitude are fixed by independently constrained
bar and gas-halo parameters rather than tuned to the data.

We do not claim that any single measurement of $R(\ell)$ will
categorically select one mechanism. The three results overlap at
small $|\ell|$, and the astrophysical uncertainty on $n_{0}^{\rm
halo}$ propagates to the estimated amplitude (the blue band of
Fig.~\ref{fig:asymmetry}), although not to the peak location. The
diagnostic information lies in the \emph{shape} of
$R(\ell)$ at $|\ell| \gtrsim 10^{\circ}$, where the three
scenarios diverge most cleanly --- the monotonic rise of an
undiluted bulge tracer, the flat DM null, and the peaked-and-declining
hadronic prediction --- and motivates dedicated
longitude-binned analyses with the full Fermi-LAT exposure.


\section{Conclusions}
\label{sec:conclusions}


We have quantified the
contribution to the Galactic Center Excess from a precessing
paraboloidal BZ jet of \SgrA{}, active $\sim 7.5\,\Myr$ ago and shut
off $\sim 2.6\,\Myr$ ago. At the EHT-favoured spin $\astar = 0.9$
the jet's hadronic cosmic rays leave an irreducible floor
of $\sim 2$--$13\%$ of the GCE surface brightness in the
$1$--$10\,\GeV$ band, rising from $\sim 2\%$ at the center to
$\sim 13\%$ at the edge of the region of interest because the
hadronic profile is flatter than the cuspy excess. In energy, the
contribution is larger: it saturates the sub-GeV data and
supplies $\sim 30$--$50\%$ of the flux in the $\sim 1$--$2\,\GeV$
range that brackets the GCE peak. The scenario does not replace the
dark-matter or millisecond-pulsar interpretations; the goal of
this work was to single out and quantify a complementary
component arising from a possible past jet of \SgrA{}, whose
earlier activity is independently established by the
Fermi/eROSITA bubbles. We would like to comment on a few aspects
of the model that delimit the scope of the present results and
point to natural extensions.

The active duration $\tact$ is the one quantity in the model that
is not fixed by an independent observation. It represents the
integrated time over which \SgrA{} sustained the MAD,
jet-launching accretion state during its past outburst, and it
enters the results principally by setting the total injected CR
energy, $E_{\CR} = \xi_{\CR} P_{\BZ}\tact$, and hence the overall
amplitude of the hadronic floor (which scales nearly linearly
with $\tact$). The physical requirement we impose is deliberately
weak: only that the jet completes more than one precession cycle,
$\Prec \equiv \tact/T_{\rm prec} > 1$, so that the azimuthal
averaging underlying the near-spherical morphology is realized.
For $T_{\rm prec} \simeq 1.5\,\Myr$ this requires merely $\tact
\gtrsim 1.5\,\Myr$; our fiducial $\tact = 7.5\,\Myr$ gives
$\Prec \simeq 5$ and sits comfortably in the allowed range.
Shorter durations ($\Prec \gtrsim$ a few) remain viable and
simply lower the floor amplitude in proportion to $\tact$; the
morphological conclusions, which depend on $\Prec > 1$ rather
than on the specific value of $\tact$, are unchanged. In this
sense $\tact$ is best read as a normalisation of the floor rather
than a tightly-constrained physical input.

The central motivation for the precessing-jet picture is not
geometric elegance but a framework to reconcile two
seemingly conflicting observations: the \SgrA{} accretion flow
(and therefore any associated jet) is significantly tilted with
respect to the Galactic rotation axis~\cite{EHT2024VIII,
Genzel2010,AnglesAlcazar2021}, while the GCE is observed to be
approximately spherical about the Galactic Center. A static
tilted jet would inject CRs into a single off-axis cone and
imprint that anisotropy on the gamma-ray sky, in tension with the
data. We have shown in detail (Sec.~\ref{ssec:injection},
Sec.~\ref{ssec:factorisation}, Appendix~\ref{app:multipole}) that
Lense--Thirring precession of the tilted disc, acting over
$\Prec > 1$ cycles, first smears the injection into an
azimuthally-symmetric band, and that isotropic diffusion through
the CMZ and bulge then homogenises this band into a
near-spherical CR cloud: the higher angular multipoles of the
injection are suppressed by the centrifugal barrier of the radial
diffusion operator, leaving a residual anisotropy at the
few-percent level. Precession is thus the mechanism that allows a
physically-motivated tilted jet to remain compatible with the
observed GCE morphology --- the tilt is an input demanded by
observation, and precession plus diffusion is what renders it
consistent with a spherical excess.

When the uniform halo is replaced by the triaxial bar-plus-halo
gas of Sec.~\ref{sec:triaxial}, the same CR transport produces a
peaked longitudinal asymmetry $I(+\ell)/I(-\ell) \simeq 1.13$
near $|\ell| \approx 10^{\circ}$, declining toward unity at
smaller and larger longitudes. This peaked shape distinguishes
the hadronic mechanism from both the null prediction of
spherically-symmetric DM and the monotonic profile of
bulge-tracing MSPs, and arises because the bar gas
{generates} the asymmetry while the symmetric gas halo
{regulates} its amplitude (Sec.~\ref{ssec:asymmetry_decomp}).
A further distinguishing feature is that the hadronic floor traces
the product $u_{\rm CR}(\bm{x}) \times n_{\rm gas}(\bm{x})$: the
hot ionized halo enters as a target-gas multiplier, whereas DM
annihilation depends on $\rho_{\rm DM}^{2}$ and MSP emission on
the bulge stellar density alone. Cross-correlating the GCE
residual with hot-halo tracers therefore offers an additional test specific to
the hadronic origin.

Our long-duration, low-luminosity outburst differs from
previous bubble-inflation scenarios in two respects that
matter for the GCE. On the one hand, the AGN-jet simulations of
Yang et al.~\cite{Yang2022} reproduce the bubbles with a
short-lived jet launched {along the Galactic rotation axis},
without tilt or precession; an axial jet is already symmetric
about the spin axis and so needs no precession to average, but by
construction it cannot accommodate the observationally required
tilt of the \SgrA{} flow. On the other hand, the model of Sarkar
et al.~\cite{Sarkar2023} inflates the bubbles with a brief,
high-power super-Eddington event from a {tilted} flow, but
again without precession; it occupies the opposite corner of the
energy--time plane from our scenario, the two delivering
comparable {total} energy through nearly inverse
combinations of power and duration. That short-burst route is
moreover disfavoured by the observed
O\,\textsc{viii}/O\,\textsc{vii} line ratio~\cite{Sarkar2023}.
Crucially for the GCE, a brief tilted burst can neither
azimuthally average the off-axis injection nor produce the
near-spherical hadronic morphology; only a duration exceeding
several precession periods allows the precession-plus-diffusion
homogenisation that the GCE morphology requires.

Several simplifications temper these conclusions. First, the
tilted, precessing disc--jet configuration is assumed to persist
coherently over the full active phase; in practice such
misaligned thick-disc systems can be subject to instabilities ---
tearing of the warped disc, disc--jet realignment via the
Bardeen--Petterson effect at the inner edge, and disruption of
coherent precession by accretion-rate fluctuations --- whose
long-term stability over several Myr is not established and would
require dedicated GRMHD modelling to verify. Second, we treat the
outburst as a single continuous episode at fixed tilt angle. A
more realistic history likely involves multiple discrete blast
episodes rather than one sustained phase, and the tilt angle
$i_{\rm tilt}$ itself may evolve as the angular-momentum
direction of the accreting gas changes over the galaxy's
history; both would alter the precise CR injection morphology and
the resulting GCE floor, and a multi-episode treatment with a
time-dependent $i_{\rm tilt}(t)$ is a natural next step. Third,
the spatial--spectral factorization and the isotropic-cloud
approximation each carry $\mathcal{O}(10$--$30\%)$ uncertainties. These limitations affect the floor
{amplitude} and the detailed morphology but not the central
qualitative result --- that a precessing tilted jet leaves an
irreducible, near-spherical hadronic imprint on the GCE. The
broader implication is specific to the GCE rather than
general: \SgrA{}'s past activity is independently recorded by the
Fermi/eROSITA bubbles, and any complete model of the GCE may
eventually have to account for the cosmic-ray reservoir that this
past fossil activity left behind.

\subsection*{Code and Data Availability}
The hadronic emissivity is
computed using the \texttt{naima} package~\cite{
Zabalza2015naima}, freely available at
\url{https://github.com/zblz/naima} with
modifications described in Sec.~\ref{ssec:numerics}. Source data
associated with the main figures of the manuscript are available
from the corresponding author upon request.

\section*{Acknowledgments}

We are grateful to Pau Amaro-Seoane for helpful feedback and a careful reading of the manuscript draft. We also thank him for useful suggestions and discussions. We additionally thank Malcolm Perry, Miguel Angel Sanchez Conde, Enrique Mier Alonso and Cristina Fernandez Suarez for discussions. The research of MJR has been funded, in part,
by the National Science Foundation under project number PHY-2309270.
MJR also wants to thank the Mitchell Family Foundation for hosting her during the Cook's Branch workshop, where
some of the research was carried out.

\appendix
\section{Magnetic Field Configuration}
\label{app:MagneticField}

\begin{figure}[!htbp]
    \centering
    \includegraphics[width=0.8\textwidth]{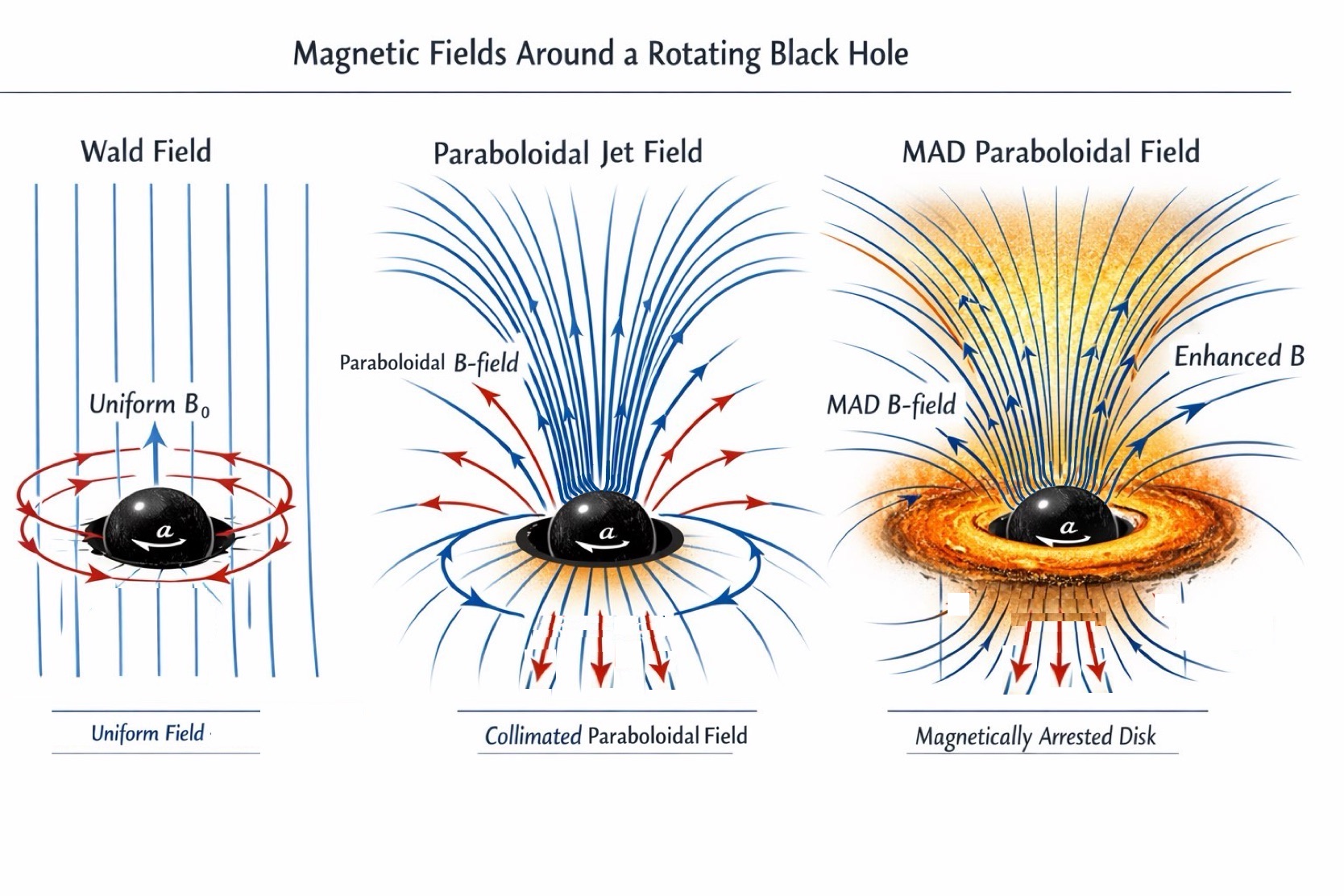}
    \caption{Comparison of electromagnetic field configurations
    around a Kerr black hole. The left panel shows the uniform
    Wald field aligned with the spin axis, the middle panel
    shows a paraboloidal field typical of jet collimation, and
    the right panel shows a paraboloidal field enhanced by a
    magnetically arrested disk (MAD), where magnetic flux
    accumulates near the horizon. Arrows indicate the direction
    of induced electric fields due to black hole rotation.}
    \label{fig:magnetic_fields}
\end{figure}

In this appendix we provide the explicit form of the vector
potential used to model the magnetosphere of \SgrA{} during
the active phase, and explain how it connects to the
BZ jet power quoted in Eq.~(\ref{eq:PBZ}) of the main text. The
model combines the paraboloidal force-free
solution~\cite{Blandford1976,McKinney2004ka,Gralla2016jets} with a
phenomenological enhancement that captures the magnetic flux
pile-up characteristic of the magnetically arrested disk (MAD)
state~\cite{Tchekhovskoy2011}.
Both observations and general-relativistic magnetohydrodynamic
(GRMHD) simulations indicate that poloidal magnetic field lines
threading accreting black holes are typically collimated along
the spin axis into an approximately paraboloidal
shape~\cite{Tchekhovskoy2008,McKinney2012MCAFs}, with additional
flux concentration near the horizon when the system saturates in
the MAD state~\cite{Tchekhovskoy2011,Narayan2012MAD}. Such
configurations are critical for collimating relativistic
outflows, guiding accelerated particles along the polar axis,
and providing the dominant energy-extraction channel via the BZ
mechanism~\cite{BZ1977} (with the magnetic Penrose
process~\cite{Tursunov2020,Dadhich2018} as a related channel). A
schematic comparison of the asymptotic uniform Wald-field,
paraboloidal, and MAD-enhanced paraboloidal configurations is shown
in Fig.~\ref{fig:magnetic_fields}; the latter is the geometry
adopted in this work.

\subsection{Vector Potential}
\label{app:Aphi}

We define the axisymmetric, stationary, vector potential
\begin{equation}
  A_{\phi}(r,\theta) = \Phi_{0}\left(\frac{r}{r_{0}}\right)^{\nu}
  \left[1 + \left(\frac{r_{+}}{r}\right)^{k}\right](1-\cos\theta),
  \label{eq:Aphi}
\end{equation}
with $A_{t} = A_{r} = A_{\theta} = 0$. The remaining
components vanish by axisymmetry ($\partial_{\phi}=0$) and by
the force-free condition.
The parameters in Eq.~(\ref{eq:Aphi}) are as follows. The
quantity $\Phi_{0}$ sets the overall flux normalization and
is fixed by the MAD saturation condition. The parameter
$r_{0}$ denotes a reference radius; we take $r_{0} = r_{+}$
such that the bracket in Eq.~(\ref{eq:Aphi}) evaluates to 2
at the horizon. The exponent $\nu$ controls the radial scaling
of the asymptotic field, $B \propto r^{\nu-2}$, and
interpolates between known force-free geometries: the
non-collimating split-monopole ($\nu = 0$)~\cite{BZ1977} and
the paraboloidal Blandford solution ($\nu = 1$), whose nested
paraboloidal surfaces collimate the outflow along the spin
axis~\cite{Blandford1976}, with the
intermediate $\nu = 3/4$ matching the time-averaged poloidal
current profile measured in GRMHD jet
simulations~\cite{Tchekhovskoy2008}. These are interpretive
limits of a single family; as our fiducial we adopt the
paraboloidal case $\nu = 1$, appropriate for a collimated jet.
Finally, the parameter $k \in [2,3]$ governs the radial extent
of the MAD enhancement; throughout this work we set $k = 2$,
consistent with GRMHD MAD simulations~\cite{Tchekhovskoy2011}.

\subsection{Angular Structure and Magnetic Components}
\label{app:Bcomponents}

The angular factor $(1-\cos\theta)$ in Eq.~(\ref{eq:Aphi})
coincides with the split-monopole / paraboloidal asymptotic
form~\cite{BZ1977,Blandford1976} and ensures $A_{\phi}=0$ on
the polar axis ($\theta=0$), as required for regularity. The
electromagnetic field tensor is defined as
\begin{equation}
F_{\mu\nu} = \partial_\mu A_\nu - \partial_\nu A_\mu .
\end{equation}
The magnetic field components are given by
\begin{equation}
B^i = \frac{1}{2}\,\epsilon^{ijk} F_{jk} .
\end{equation}
The corresponding magnetic field components associated with
\eqref{eq:Aphi} are
\begin{align}
  B^{r}(r,\theta)
  &= \frac{1}{\sqrt{-g}}\,\partial_{\theta}A_{\phi}
  \;=\; \frac{\Phi_{0}}{\sqrt{-g}}\,
        \left(\frac{r}{r_{0}}\right)^{\nu}
        \left[1+\left(\frac{r_{+}}{r}\right)^{k}\right]\sin\theta,
        \label{eq:Br}\\[4pt]
  B^{\theta}(r,\theta)
  &= -\frac{1}{\sqrt{-g}}\,\partial_{r}A_{\phi},
        \label{eq:Btheta}
\end{align}
where $\sqrt{-g} = (r^{2}+a^{2}\cos^{2}\theta)\sin\theta$ on
the Kerr metric. At the horizon $r = r_{+}$, the radial field
reduces to
\begin{equation}
  B^{r}(r_{+},\theta) = \frac{2\,\Phi_{0}}{r_{+}^{2}+a^{2}
                              \cos^{2}\theta},
  \label{eq:Bhorizon}
\end{equation}
which is mildly bunched toward the pole through the
$a^{2}\cos^{2}\theta$ term in the denominator
(cf.~\cite{Gralla2016jets}). At large radii $r \gg r_{+}$ the
bracket in Eq.~(\ref{eq:Br}) reduces to unity and the field
smoothly connects to the asymptotic paraboloidal configuration
$B \propto r^{\nu - 2}$.

\subsection{Connection to the BZ Power}
\label{app:BZ}

The BZ jet power follows directly from $\Phi_{\mathrm{BH}}$
 and the horizon angular velocity $\Omega_{H} = \astar c / (2 r_{+})$  via \eqref{eq:PBZ1}
where $\kappa$ is a dimensionless geometry coefficient
determined by the angular structure of the field lines. For
the asymptotic paraboloidal geometry encoded in
Eq.~(\ref{eq:Aphi}), $\kappa$ can be computed analytically
from the integral $\kappa = K/(4\pi^{2})$ with $K =
2\int_{0}^{1}\tilde{I}(x)\,\tilde{\Omega}(x)\,dx$, where
$\tilde{I}(x)$ and $\tilde{\Omega}(x)$ are the dimensionless
current and angular-velocity functions of the paraboloidal
configuration~\cite{Gralla2016jets}. Numerical evaluation
yields
\begin{equation}
  \kappa_{\mathrm{par}} \simeq 0.0445,
  \label{eq:kappa}
\end{equation}
in agreement with the value $\kappa \approx 0.044$ extracted
from GRMHD simulations of MAD jets.
The MAD enhancement factor $[1+(r_{+}/r)^{k}]$ in
Eq.~(\ref{eq:Aphi}) modifies the radial profile of the field
but leaves $\kappa$ essentially unchanged, since the latter
depends only on the angular structure of $A_{\phi}$ on the
horizon and at infinity, both of which remain
$(1-\cos\theta)$. The role of the enhancement is therefore to
ensure a physically reasonable $B(r)$ falloff for downstream
cosmic-ray injection, not to alter the BZ luminosity itself.

\subsection{Validity of the Force-Free Approximation}
\label{app:FFE}

The vector potential in Eq.~(\ref{eq:Aphi}) is a force-free
configuration in the limit of negligible plasma inertia
($\sigma \equiv B^{2}/4\pi\rho c^{2} \gg 1$). Inside the
funnel of a MAD black-hole magnetosphere this condition is
satisfied ($\sigma \gg 10^{2}$ from GRMHD
simulations~\cite{McKinney2012MCAFs}), justifying the use of
force-free electrodynamics for the BZ-power estimate. Outside
the funnel --- in the equatorial accretion flow --- the
force-free approximation breaks down, but this region does
not contribute to the jet luminosity at leading order.
Reconnection events at the equatorial current sheet, expected
to occur intermittently in the MAD
state~\cite{Ripperda2022}, are not included in our model and
may provide an additional source of non-thermal particles.

\section{Multipole analysis of the diffusion equation}
\label{app:multipole}

This appendix quantifies the diffusive isotropisation argument given
in Sec.~\ref{ssec:factorisation} and justifies the monopole-only
treatment used in the main text. We expand the CR density in
spherical multipoles, derive closed-form expressions for both the
geometric source coefficients and the diffusive damping of each
multipole, and verify the residual anisotropy numerically.

\subsection{The diffusion equation}
\label{app:multipole:eq}

In the bulge interior the CR number density at fixed energy
satisfies the time-dependent isotropic diffusion equation
\begin{equation}
  \frac{\partial n_{\CR}(\bm{r},t)}{\partial t}
  \;=\; D\,\nabla^{2}\,n_{\CR}(\bm{r},t)
        \;+\; \mathcal{S}(\bm{r},t)
  \label{eq:diffusion_eq}
\end{equation}
where $\nabla^{2}$ is the 3D Laplacian in spherical coordinates,
$\mathcal{S}(\bm{r},t)$ is the BZ-jet CR injection rate per unit
volume
, and
$D$ is the (energy-dependent) diffusion coefficient. Outside the
multipole analysis below we use a two-zone $D(r)$ with $D_{\rm CMZ}$
for $r<R_{b}$ and $D_{\rm bulge}$ for $r>R_{b}$, as described in
Sec.~\ref{ssec:Green}; in this appendix we take a single
$D = D_{\rm bulge}$ for analytical tractability.

For the source we adopt the precession-averaged BZ-cone pattern of
Eq.~(\ref{eq:Ptot}), localized on a shell at $r_{\rm dep}$,
\begin{equation}
  \mathcal{S}(\bm{r},t)
  \;=\; \mathcal{S}_{0}\,
        P(\vartheta)\,
        \delta(r - r_{\rm dep})\,
        \Theta(\tact - t),
  \label{eq:source_shell}
\end{equation}
with normalisation $\mathcal{S}_{0}$ fixed by total injected CR
energy and the Heaviside $\Theta$ encoding the active phase.
\subsection{Source decomposition}
\label{app:multipole:source}
The precession-averaged injection pattern $ P(\vartheta)$ of
Eq.~(\ref{eq:Ptot}) is azimuthally symmetric about the Galactic
rotation axis and invariant under $\vartheta\to\pi-\vartheta$, so it
expands in even-$\ell$ Legendre polynomials,
\begin{equation}
  P(\vartheta)\;=\;
    \sum_{\ell\ge 0}\,a_{\ell}\,P_{\ell}(\cos\vartheta),
  \qquad
  a_{\ell} \;=\;
    \frac{2\ell+1}{2}\int_{0}^{\pi}\!
        P(\vartheta)\,
        P_{\ell}(\cos\vartheta)\,\sin\vartheta\,d\vartheta,
  \label{eq:Pl_coefs}
\end{equation}
with $a_{\ell} = 0$ for odd $\ell$ by the bipolar symmetry
$P(\vartheta) = P(\pi - \vartheta)$, which forces all odd-parity
Legendre projections to vanish. For non-overlapping cones ($i_{\tilt} + \theta_{c} < \pi/2$,
satisfied at the fiducial $35^{\circ}+20^{\circ}=55^{\circ}$), the
integral can be evaluated analytically. We outline the three steps.

\paragraph{Step 1: rewrite $P(\vartheta)$ as a cap indicator.} 
The precession-averaged $P(\vartheta)$ has a closed-form
expression involving $\arccos$, but plugging this into the Legendre
projection yields an integral with a transcendental weight that does
not evaluate in elementary form. The cap-indicator representation
writes $P(\vartheta)$ instead as a time-average
of a binary $\{0,1\}$ indicator function over the precession cycle.
Although mathematically equivalent, this representation lets us swap
the order of integration: the Legendre projection becomes a
two-dimensional integral over a spherical cap, which evaluates
analytically via the addition theorem. The conceptual move is to
trade "compute $P$, then project" for "use $P$'s construction as an
average of caps, project each cap, then average." Only the second
admits a closed-form result.

The precession-averaged injection pattern $P(\vartheta)$ admits a
useful integral representation as a time-average over the precession
cycle of the indicator function for the BZ-jet cone. At any
precession phase $\varphi_{\rm jet} \in [0, 2\pi)$, the jet axis
points at $(i_{\tilt}, \varphi_{\rm jet})$ on the celestial sphere
and illuminates a spherical cap of half-angle $\theta_{c}$ around
that direction. A given observation direction $(\vartheta, 0)$
(set to azimuth $0$ without loss of generality, since $P$ is
axisymmetric) lies inside the cap when the angular distance
$\Delta(\vartheta, \varphi_{\rm jet})$ between
$(\vartheta, 0)$ and $(i_{\tilt}, \varphi_{\rm jet})$ satisfies
$\Delta < \theta_{c}$. Time-averaging over the precession cycle
yields
\begin{equation}
  P(\vartheta) \;=\; \frac{1}{2\pi}\int_{0}^{2\pi}\!d\varphi_{\rm jet}\,
                     H\bigl(\cos\theta_{c} -
                            \cos\Delta(\vartheta,\varphi_{\rm jet})\bigr),
  \label{eq:P_heaviside}
\end{equation}
where $H(\cdot)$ is the Heaviside step function. The argument of
$H$ is positive when $\Delta < \theta_{c}$ (the observation
direction is inside the cap, $H = 1$) and negative when
$\Delta > \theta_{c}$ (outside the cap, $H = 0$). The angular
distance $\Delta$ is given by the spherical law of cosines,
\begin{equation}
  \cos\Delta(\vartheta, \varphi_{\rm jet})
  \;=\; \cos i_{\tilt}\,\cos\vartheta
        \;+\; \sin i_{\tilt}\,\sin\vartheta\,\cos\varphi_{\rm jet}\,,
  \label{eq:Delta_def}
\end{equation}

We introduce two angles to locate the jet axis on the celestial sphere:
the {tilt angle} $i_{\rm tilt}$ between the jet axis and the
Galactic rotation axis (fixed, $35^\circ$ fiducial), and the
{precession phase} $\varphi_{\rm jet}$, the azimuthal angle of
the jet axis around the Galactic axis ($\varphi_{\rm jet}$ varies from
$0$ to $2\pi$ over one precession period as the jet axis traces out a
cone in 3D). At any instant the jet illuminates a spherical cap of
half-angle $\theta_c$ centered on $(i_{\rm tilt}, \varphi_{\rm jet})$;
the time-averaged pattern $P(\vartheta)$ is obtained by averaging this
illumination over $\varphi_{\rm jet}\in[0,2\pi)$.
Substituting the integral representation Eq.~(\ref{eq:P_heaviside})
for $P(\vartheta)$ into the Legendre projection
Eq.~(\ref{eq:Pl_coefs}), using the axisymmetry of $P$ to lift the
1D angular integral to a 2D integral over $\Omega$ (which
introduces an additional factor of $1/(2\pi)$), and noting that the
Heaviside restricts the $\Omega$ integration to the
spherical cap, gives
\begin{equation}
  a_{\ell} \;=\;
  \frac{2\ell+1}{8\pi^{2}}
  \int_{0}^{2\pi}\!d\varphi_{\rm jet}
  \int_{\mathrm{cap}(i_{\tilt}, \varphi_{\rm jet})}\!
    P_{\ell}(\cos\vartheta)\,d\Omega.
  \label{eq:al_cap}
\end{equation}

\paragraph{Step 2: evaluate the cap integral via the addition theorem.}
The integrand $P_{\ell}(\cos\vartheta)$ in Eq.~(\ref{eq:al_cap}) is
expressed in coordinates aligned with the Galactic axis (polar angle
$\vartheta$), but the cap is naturally parameterized in
cap-centered coordinates: a polar angle $\psi$ measured from
the cap center (with $\psi \le \theta_{c}$ inside the cap) and an
azimuth $\chi$ around the cap axis. The two frames are related by a
rotation through $i_{\tilt}$, and the spherical-harmonic addition
theorem expresses $P_{\ell}(\cos\vartheta)$ in cap-centered
coordinates:
\begin{equation}
  P_{\ell}(\cos\vartheta) \;=\;
    P_{\ell}(\cos i_{\tilt})\,P_{\ell}(\cos\psi)
    \;+\; 2\sum_{m=1}^{\ell}
      \frac{(\ell-m)!}{(\ell+m)!}\,
      P_{\ell}^{m}(\cos i_{\tilt})\,
      P_{\ell}^{m}(\cos\psi)\,\cos(m\chi)\,,
  \label{eq:addition_theorem}
\end{equation}
where $P_{\ell}^{m}$ are the associated Legendre functions. The cap
is rotationally symmetric about its own axis ($\chi \in [0,2\pi]$),
so integrating $\cos(m\chi)$ over a full period kills every $m\ge 1$
term. Only the $m=0$ projection survives:
\begin{equation}
  \int_{\rm cap}P_{\ell}(\cos\vartheta)\,d\Omega
  \;=\; 2\pi\,P_{\ell}(\cos i_{\tilt})\,
        \int_{0}^{\theta_{c}}\!P_{\ell}(\cos\psi)\,\sin\psi\,d\psi
  \;=\; 2\pi\,P_{\ell}(\cos i_{\tilt})\,
        \int_{\cos\theta_{c}}^{1}\!P_{\ell}(u)\,du.
  \label{eq:cap_integral}
\end{equation}
The two factors carry distinct physical content:
$P_{\ell}(\cos i_{\tilt})$ encodes the {orientation} of the
cap with respect to the Galactic axis (a rotation-matrix element),
while $\int_{\cos\theta_{c}}^{1}P_{\ell}(u)\,du$ encodes the
{shape} of the cap (its angular extent).

\paragraph{Step 3: add the conjugate cone and evaluate the radial integral.}
The bipolar BZ source illuminates both an upper cone (centered at
polar angle $i_{\tilt}$) and a lower cone (centered at $\pi -
i_{\tilt}$). Using
$P_{\ell}(\cos(\pi - i_{\tilt})) = (-1)^{\ell}\,P_{\ell}(\cos i_{\tilt})$,
adding the two cone contributions gives a factor $[1 + (-1)^{\ell}]$,
which vanishes for odd $\ell$ and equals 2 for even $\ell$ ---
recovering the bipolar-symmetry parity selection. 

A useful form is
\begin{equation}
  (2\ell+1)\,P_{\ell}(u)
  \;=\; \frac{d}{du}\bigl[P_{\ell+1}(u) - P_{\ell-1}(u)\bigr].
  \label{eq:legendre_derivative}
\end{equation}
allows integrating Eq.~(\ref{eq:legendre_derivative}) in closed
form,
\begin{equation}
  (2\ell+1) \int_{\cos\theta_{c}}^{1}\! P_{\ell}(u)\,du
  \;=\; \Bigl[P_{\ell+1}(u) - P_{\ell-1}(u)\Bigr]_{\cos\theta_{c}}^{1}
  \end{equation}
  \begin{equation}
  = P_{\ell-1}(\cos\theta_{c}) - P_{\ell+1}(\cos\theta_{c})
  \;=\;
  \frac{2\ell+1}{\ell(\ell+1)}\,(1 - \cos^{2}\theta_{c})\,
       P'_{\ell}(\cos\theta_{c})
 \label{eq:u_integral_evaluated}
\end{equation}

Combined
with $\sin^{2}\theta_{c}/(1-\cos\theta_{c}) = 1+\cos\theta_{c}$, the
ratio 
\begin{equation}
  \boxed{\;
    \frac{a_{\ell}}{a_{0}}
    \;=\; \frac{2\ell+1}{\ell(\ell+1)}\,
          \bigl(1 + \cos\theta_{c}\bigr)\,
          P_{\ell}(\cos i_{\tilt})\,
          P'_{\ell}(\cos\theta_{c}),
    \qquad \ell\ \text{even},\ \ell \ge 2,
  \;}
  \label{eq:a_l_compact}
\end{equation}
with $a_{0} = 1 - \cos\theta_{c}$. The tilt and cone-width
dependences factorise: $P_{\ell}(\cos i_{\tilt})$ controls the
tilt dependence, and $(1+\cos\theta_{c})\,P'_{\ell}(\cos\theta_{c})$
the cone-width dependence. Consequently $a_{\ell}$ vanishes at the
zeros of $P_{\ell}(\cos i_{\tilt})$.
For our fiducial $i_{\tilt}=35^{\circ}$, $\theta_{c}=20^{\circ}$,
Eq.~(\ref{eq:a_l_compact}) gives $a_{0}\simeq 0.060$,
$a_{2}\simeq 0.139$, $a_{4}\simeq -0.067$, $a_{6}\simeq -0.156$,
in agreement with direct numerical Legendre projection of
Eq.~(\ref{eq:Pl_coefs}).

\subsection{Multipole structure of the CR density}
\label{app:multipole:structure}

Because Eq.~(\ref{eq:diffusion_eq}) separates in spherical
coordinates and the source~(\ref{eq:source_shell}) is axisymmetric
about the spin axis, the resulting CR density admits the multipole
decomposition
\begin{equation}
  n_{\CR}(r,\vartheta,t) \;=\;
    \sum_{\ell\ge 0}\, n_{\ell}(r,t)\,
        P_{\ell}(\cos\vartheta),
  \qquad
  n_{\ell}(r,t) \;\equiv\;
    a_{\ell}\,b_{\ell}(r,t),
  \label{eq:nCR_multipole}
\end{equation}
where $b_{\ell}(r,t)$ is the radial Green function for mode $\ell$,
satisfying the radial form of Eq.~(\ref{eq:diffusion_eq}):
\begin{equation}
  \frac{\partial b_{\ell}}{\partial t}
  \;=\; D\,\Biggl[\frac{1}{r^{2}}\,\frac{\partial}{\partial r}
        \!\left(r^{2}\,\frac{\partial b_{\ell}}{\partial r}\right)
        - \frac{\ell(\ell+1)}{r^{2}}\,b_{\ell}\Biggr],
  \label{eq:radial_diff}
\end{equation}
with the centrifugal term $\ell(\ell+1)/r^{2}$ providing the
$\ell$-dependent suppression that drives diffusive isotropisation.
The monopole $n_{0}(r,t)\equiv a_{0}\,b_{0}(r,t)$ is the spherically
averaged CR density --- the only term retained in the main-text
approximation $n_{\CR}(\bm{x})\approx n_{\CR}(r)\equiv n_{0}(r,t)$.
The higher moments $n_{2}, n_{4},\ldots$ are the residual
anisotropies bounded below. The multipole-to-monopole ratio
\begin{equation}
  \frac{n_{\ell}(r,t)}{n_{0}(r,t)}
  \;=\; \frac{a_{\ell}}{a_{0}}\,
        \frac{b_{\ell}(r,t)}{b_{0}(r,t)}
  \label{eq:nl_n0_full}
\end{equation}
factorises into the source-geometry ratio from
Eq.~(\ref{eq:a_l_compact}) and the diffusive damping ratio derived
in the next subsection.

\subsection{Analytical solution: point source at the origin}
\label{app:multipole:analytic}

In the limit where the source size is much smaller than the
diffusion length, $r_{\rm dep}\ll\lambda_{\rm bulge}\equiv\sqrt{4 D\tact}$,
the shell at $r_{\rm dep}$ acts as an effective point source at the
origin. Solving Eq.~(\ref{eq:radial_diff}) with a delta initial
condition, the Hankel--Fourier representation
\begin{equation}
  b_{\ell}(r,t) \;\propto\; \int_{0}^{\infty}\!dk\,k^{2}\,
                     j_{\ell}(kr)\,e^{-D k^{2} t}
  \label{eq:bl_hankel}
\end{equation}
evaluates in closed form by Gradshteyn--Ryzhik \cite{GradshteynRyzhik} Eq.~6.631.4. Using
Kummer's transformation
${}_{1}F_{1}(a;b;z) = e^{z}\,{}_{1}F_{1}(b-a;b;-z)$ to remove the
Gaussian envelope from the hypergeometric argument, the ratio of
mode $\ell$ to monopole reads
\begin{equation}
  \boxed{\;
    \frac{b_{\ell}(r,t)}{b_{0}(r,t)}
    \;=\; \frac{\Gamma\!\left(\frac{\ell+3}{2}\right)}
                {\Gamma\!\left(\ell+\frac{3}{2}\right)}\,
          \Biggl(\frac{r}{2\sqrt{D t}}\Biggr)^{\!\ell}\,
          {}_{1}F_{1}\!\left(\frac{\ell}{2};\;\ell+\frac{3}{2};\;
                              \frac{r^{2}}{4 D t}\right).
  \;}
  \label{eq:bl_closedform}
\end{equation}
Two limits are useful. For $r\ll\sqrt{D t}$, ${}_{1}F_{1}\to 1$ and
the damping reduces to a power law,
\begin{equation}
  \frac{b_{\ell}}{b_{0}} \;\approx\;
       \frac{\Gamma\!\left(\frac{\ell+3}{2}\right)}
            {\Gamma\!\left(\ell+\frac{3}{2}\right)}\,
       \Biggl(\frac{r}{2\sqrt{D t}}\Biggr)^{\!\ell},
  \label{eq:bl_leading}
\end{equation}
so a delta source at the origin produces no angular structure at
$r\to 0$, as required by regularity. For $r\gg\sqrt{D t}$,
$b_{\ell}/b_{0}\to 1$ and the geometric source pattern is preserved.
Eq.~(\ref{eq:bl_closedform}) interpolates smoothly between these
regimes and satisfies Eq.~(\ref{eq:radial_diff}) exactly, verified
to machine precision for $\ell\in\{0,2,4,6\}$.

We are not aware of a previous application of
Eq.~(\ref{eq:bl_closedform}) to anisotropic cosmic-ray sources. The
underlying mathematical-physics expansion is standard
\cite{Watson1944},\cite{MorseFeshbach2004}, but its use as a
diagnostic for diffusive isotropisation of an injected angular
pattern appears to be new to this work.

\subsection{Shell source}
\label{app:multipole:shell}

The physically correct boundary condition for our model is the
shell source~(\ref{eq:source_shell}) at $r_{\rm dep}$ with angular
weight $P(\vartheta)$, emitting continuously over $[0,\tact]$. The
instantaneous Green function for mode $\ell$, obtained from the
addition theorem of the 3D Gaussian heat kernel in modified
spherical Bessel functions, is
\begin{equation}
  Q_{\ell}(r,r',\tau)
  \;=\; \frac{1}{(4\pi D\tau)^{3/2}}\,
        \exp\!\left[-\frac{r^{2}+r'^{2}}{4 D\tau}\right]\,
        i_{\ell}\!\left(\frac{r\,r'}{2 D\tau}\right),
  \label{eq:Ql_shell}
\end{equation}
where $i_{\ell}(z) \equiv \sqrt{\pi/(2z)}\,I_{\ell+1/2}(z)$ is the
modified spherical Bessel function of the first kind, regular at
the origin. For continuous injection over the active phase, the
damping ratio is the time-integrated form
\begin{equation}
  \boxed{\;
    \frac{b_{\ell}(r,\tact)}{b_{0}(r,\tact)}
    \;=\; \frac{\int_{0}^{\tact}\!d\tau\,Q_{\ell}(r,r_{\rm dep},\tau)}
                {\int_{0}^{\tact}\!d\tau\,Q_{0}(r,r_{\rm dep},\tau)}\,.
  \;}
  \label{eq:bl_shell_cont}
\end{equation}
This is the formula used to compute the table below. For
$r_{\rm dep}\to 0$, $i_{\ell}(rr'/(2D\tau))\to (rr'/2D\tau)^{\ell}/(2\ell+1)!!$,
and Eq.~(\ref{eq:bl_shell_cont}) reduces to the point-source
expression~(\ref{eq:bl_closedform}) up to a normalisation that
cancels in the ratio. For our fiducial parameters
$r_{\rm dep}/\lambda_{\rm bulge}\simeq 0.06$, the two formulations
agree to leading order, with corrections of order
$(r_{\rm dep}/\lambda_{\rm bulge})^{2}\sim 4\times 10^{-3}$.

\subsection{Numerical results}
\label{app:multipole:numerical}

Combining Eq.~(\ref{eq:nl_n0_full}) with the source coefficients
from Eq.~(\ref{eq:a_l_compact}) and the damping factors from
Eq.~(\ref{eq:bl_shell_cont}) at fiducial $\tact = 7.5\,\Myr$,
$D_{\rm bulge} = 3\times 10^{28}\,\mathrm{cm^{2}\,s^{-1}}$,
$r_{\rm dep} = 100\,\mathrm{pc}$ gives the multipole ratios listed
in Table~\ref{tab:multipole_ratios}. These have been cross-checked
against direct 4D integration of the time-integrated Gaussian Green
function over the source shell, followed by Legendre projection at
the observer; both methods agree at the level of numerical-quadrature
precision.

\begin{table}[!htbp]
\centering
\begin{tabular}{lccccc}
\hline
$r$ (kpc)             & $0.5$    & $1.0$    & $1.5$    & $2.0$    & $3.0$    \\
\hline
$n_{2}(r)/n_{0}(r)$   & $+0.027$ & $+0.011$ & $+0.009$ & $+0.009$ & $+0.011$ \\
$n_{4}(r)/n_{0}(r)$   & $-3\times 10^{-4}$
                      & $-3\times 10^{-5}$
                      & $-1\times 10^{-5}$
                      & $-1\times 10^{-5}$
                      & $-1\times 10^{-5}$ \\
\hline
\end{tabular}
\caption{Multipole-to-monopole ratios from
Eqs.~(\ref{eq:nl_n0_full})--(\ref{eq:bl_shell_cont}) at fiducial
source pattern and diffusion parameters. The quadrupole sits at a
few percent and is nearly $r$-independent. The hexadecapole is
negative (because the source coefficient $a_{4}$ is negative) and
an additional factor $\sim 10^{2}$ smaller than the quadrupole;
higher even multipoles are below $10^{-5}$.}
\label{tab:multipole_ratios}
\end{table}

The observed surface brightness is a LOS integral weighted toward
$r\sim 0.5$--$1\,\kpc$ in the inner ROI ($\theta\lesssim 6^{\circ}$,
impact parameter $\rho\lesssim 0.86\,\kpc$), so the LOS-weighted
quadrupole-to-monopole ratio inherits the bound
\begin{equation}
  \bigl|\langle n_{2}\rangle_{\rm LOS}
       \big/\langle n_{0}\rangle_{\rm LOS}\bigr|
  \;\lesssim\; 3\times 10^{-2}\,.
  \label{eq:LOS_bound}
\end{equation}
The CR cloud is therefore monopolar at the $\sim 3\%$ level or
better, validating the isotropic approximation
$n_{\CR}(\bm{x})\approx n_{0}(r,t)$ used in the main text. The
associated angular systematic on the GCE brightness profile is well
below the $\sim 10\%$ statistical errors on the data of
\cite{DiMauro2021}, and is subsumed within the factorisation budget
of Sec.~\ref{ssec:factorisation}.

\section{Closed-form quasi-stationary two-zone solution}
\label{app:greenbox}

The fiducial radial profile $u_{\CR}(r)$ is obtained
numerically (Sec.~\ref{ssec:numerics}). It is nonetheless
useful to have a closed-form solution, both to expose the
parametric dependence of the profile and to provide an
independent check on the numerics. Here we construct the
quasi-stationary matched solution and compare it against the
direct numerical evolution.

\paragraph{Single-zone Green function.}
For a single-zone medium with constant $D$ and a continuous
spherically-symmetric point source active over $[0, T]$, the
time-integrated Green-function solution~\cite{Atoyan1995} is
\begin{equation}
  u_{\CR}(r, t{=}T) =
  \frac{Q_{0}}{4\pi D\,r}\,
  \erfc\!\left(\frac{r}{\lambda}\right),
  \qquad \lambda \equiv \sqrt{4\,D\,T},
  \label{eq:Green_single}
\end{equation}
evaluated at the end of the active phase. The $1/r$ behaviour
is the point-source singularity, regulated by $r_{\rm dep}$ in
the LOS integral. The length $\lambda \equiv \sqrt{4DT}$ is
the Green-function diffusion scale; it differs by an
$\mathcal{O}(1)$ coefficient from the $\sqrt{6DT}$
mean-square-displacement scale used for the escape time in
Sec.~\ref{ssec:spectrum}, the two being distinct conventional
measures of the same diffusion length.

\paragraph{Matched two-zone solution.}
With the two-zone diffusivity Eq.~(\ref{eq:2zone}), continuity
of $u_{\CR}$ and of the diffusive flux
$D\,\partial_r u_{\CR}$ at $r = R_b$ determines the matched
solution. The interface conditions are
\begin{equation}
  u_{\CR}^{\mathrm{in}}(R_{b}) =
  u_{\CR}^{\mathrm{out}}(R_{b}),
  \qquad
  D_{\mathrm{CMZ}}\,
  \partial_{r} u_{\CR}^{\mathrm{in}}\big|_{R_{b}} =
  D_{\mathrm{bulge}}\,
  \partial_{r} u_{\CR}^{\mathrm{out}}\big|_{R_{b}},
  \label{eq:interface}
\end{equation}
yielding
\begin{equation}
  u_{\CR}(r) =
  \begin{cases}
    \dfrac{Q_{0}}{4\pi D_{\mathrm{CMZ}}\,r}\,
      \erfc\!\left(\dfrac{r}{\lambda_{\mathrm{CMZ}}}\right)
      + \dfrac{b_{2}}{r}, & r < R_{b},\\[10pt]
    \dfrac{b_{1}\,Q_{0}}{4\pi D_{\mathrm{bulge}}\,r}\,
      \erfc\!\left(\dfrac{r}{\lambda_{\mathrm{bulge}}}\right),
      & r \ge R_{b},
  \end{cases}
  \label{eq:GreenBox}
\end{equation}
with
\begin{align}
  b_{1} &= \frac{2\,e^{-R_{b}^{2}/\lambda_{\mathrm{CMZ}}^{2}}}{\sqrt{\pi}\,\lambda_{\mathrm{CMZ}}}
          \left[
            \frac{2\,e^{-R_{b}^{2}/\lambda_{\mathrm{bulge}}^{2}}}
                 {\sqrt{\pi}\,\lambda_{\mathrm{bulge}}}
            + \frac{\erfc(R_{b}/\lambda_{\mathrm{bulge}})}{R_{b}}
              \left( 1 - \frac{D_{\mathrm{CMZ}}}{D_{\mathrm{bulge}}} \right)
          \right]^{-1},
  \label{eq:b1} \\[4pt]
  b_{2} &= \frac{Q_{0}}{4\pi}\!\left[
            \frac{b_{1}\,\erfc(R_{b}/\lambda_{\mathrm{bulge}})}
                 {D_{\mathrm{bulge}}}
            - \frac{\erfc(R_{b}/\lambda_{\mathrm{CMZ}})}
                   {D_{\mathrm{CMZ}}}
          \right].
  \label{eq:b2}
\end{align}
For the fiducial parameters (broadened present-day
$\lambda$), $b_1 \simeq 0.47$. We have verified that
Eq.~(\ref{eq:b1}) satisfies both interface conditions of
Eq.~(\ref{eq:interface}) to machine precision and that it
reduces correctly to the single-zone form
Eq.~(\ref{eq:Green_single}), $b_1\to1$, $b_2\to0$ in the
limit $D_{\mathrm{CMZ}}\to D_{\mathrm{bulge}}$.

\begin{figure}[t]
  \centering
  \includegraphics[width=0.6\textwidth]{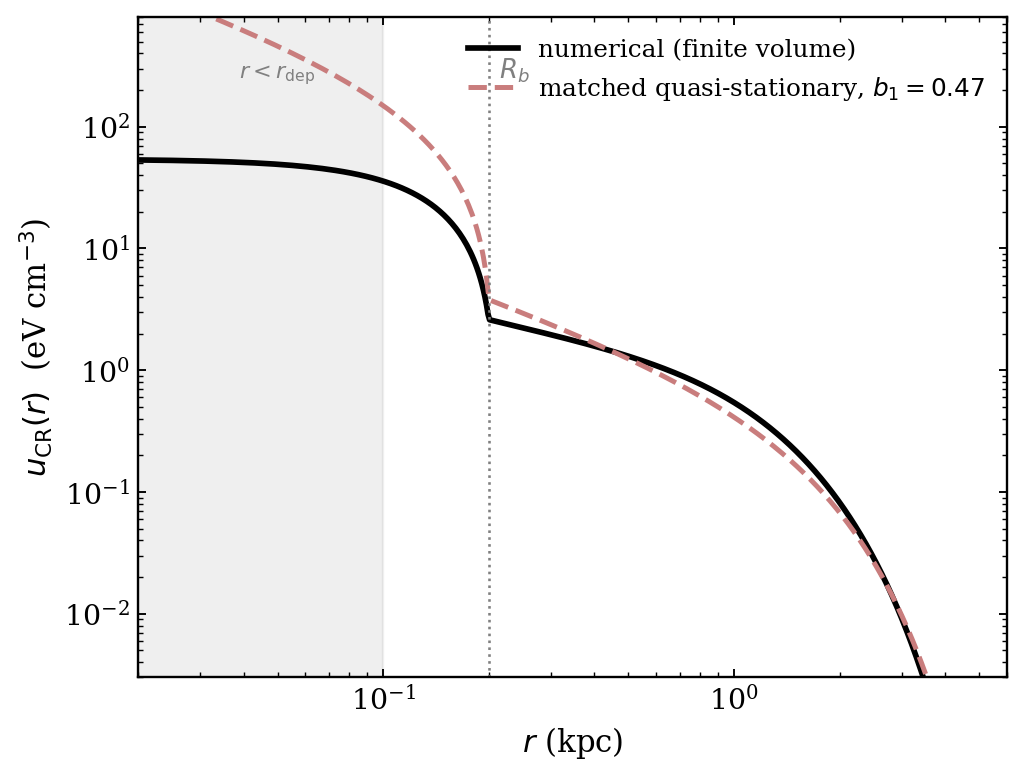}
    \caption{Present-day CR energy density from the two-zone
  transport problem, Eq.~(\ref{eq:diffusion_pde}): direct
  finite-volume solution (solid) versus the matched
  quasi-stationary form Eq.~(\ref{eq:GreenBox}) with
  $b_1 = 0.47$ (dashed), both normalized to
  $E_{\CR} = 6.2\times10^{53}\,\erg$. The
  analytic form agrees to $\lesssim 25\%$ over the
  GCE-dominant radii ($0.4 \lesssim r \lesssim 3\,\kpc$) but
  over-confines the inner zone (factor $\sim 4$ at
  $100\,\mathrm{pc}$), where the quasi-stationary assumption
  fails because the CMZ crossing time
  $R_b^2/6D_{\rm CMZ}\simeq2\,\Myr$ is short compared to
  $\tact$. The shaded region $r < r_{\rm dep}$ is excluded from
  the line-of-sight integral.}
  \label{fig:greenbox}
\end{figure}

\paragraph{Range of validity.}
Equation~(\ref{eq:GreenBox}) is the matched solution of the
{stationary} two-zone problem; treating the active-phase
profile as quasi-stationary is the only approximation,
controlled by $\lambda_{\rm CMZ}, R_b \ll \lambda_{\rm bulge}$.
Because the CMZ crossing time
$R_b^2/6D_{\rm CMZ}\simeq2\,\Myr$ is short compared to $\tact$,
the inner zone fills and, after shutoff, partially drains ---
dynamics a frozen quasi-stationary profile cannot represent ---
so the closed form over-confines the inner zone. Figure
\ref{fig:greenbox} compares Eq.~(\ref{eq:GreenBox}) with the
direct numerical solution of Eq.~(\ref{eq:diffusion_pde}):
the two agree to $\lesssim 25\%$ over the GCE-dominant radii
$0.4 \lesssim r \lesssim 3\,\kpc$, while the quasi-stationary
form exceeds the numerical profile by a factor $\sim 1.5$ at
$R_b$ and $\sim 4$ at $100\,\mathrm{pc}$. Since the LOS
integral excludes $r < r_{\rm dep} = 100\,\mathrm{pc}$ and is
dominated by the agreement region, the closed form serves as a
useful analytic guide and validation, but the fiducial results
use the numerical profile throughout.

\bibliographystyle{unsrt}
\bibliography{biblioBZ}

\end{document}